\documentclass[pdflatex,sn-mathphys-num]{sn-jnl}% Math and Physical Sciences Numbered Reference Style
\usepackage{graphicx}%
\usepackage{multirow}%
\usepackage{amsmath,amssymb,amsfonts}%
\usepackage{amsthm}%
\usepackage{mathrsfs}%
\usepackage[title]{appendix}%
\usepackage{xcolor}%
\usepackage{textcomp}%
\usepackage{manyfoot}%
\usepackage{booktabs}%
\usepackage{algorithm}%
\usepackage{algorithmicx}%
\usepackage{algpseudocode}%
\usepackage{listings}%
\usepackage{siunitx}
\usepackage{caption}
\usepackage{pdfpages}
\theoremstyle{thmstyleone}%
\theoremstyle{thmstyletwo}%

\theoremstyle{thmstylethree}%

\usepackage{titlesec}
\titleformat{\section}{\normalfont\large\bfseries}{ }{0pt}{}
\titleformat{\subsection}{\normalfont\normalsize\bfseries}{ }{0pt}{}

\begin{document}

%\title[Article Title]{Fe-catalysed on-surface synthesis of purely covalent 2D Graphdiyne Networks} 
\title[Article Title]{Iron-mediated on-surface synthesis of substrate-decoupled graphdiyne monolayers} 

\author[1]{\fnm{Alice} \sur{Cartoceti}}%\email{alice.cartoceti@polimi.it}
\equalcont{These authors contributed equally to this work.}

\author*[2,3]{\fnm{Simona} \sur{Achilli}}\email{simona.achilli@unimi.it}
\equalcont{These authors contributed equally to this work.}

\author[1]{\fnm{Gianni} \sur{Conti}}
%\author[2]{\fnm{Masoumeh} \sur{Alihosseini}}

\author[4,7]{\fnm{Eliecer} \sur{Pelaez-Sifonte}}

\author[5]{\fnm{Alessio} \sur{Orbelli Biroli}}%\email{alessio.orbellibiroli@unipv.it}

\author[6]{\fnm{Sabine} \sur{Maier}}

\author[7]{\fnm{Francesco} \sur{Sedona}}%\email{francesco.sedona@unipd.it}

\author[1]{\fnm{Paolo} \sur{D'Agosta}}

\author[1]{\fnm{Francesco} \sur{Tumino}}

\author[1]{\fnm{Andrea} \sur{Li Bassi}}%\email{andrea.libassi@polimi.it}

\author*[4,8]{\fnm{Jorge} \sur{Lobo-Checa}}\email{jorge.lobo@csic.es}

\author*[1]{\fnm{Carlo S.} \sur{Casari}}\email{carlo.casari@polimi.it}

\affil*[1]{\orgdiv{Department of Energy}, \orgname{Politecnico di Milano}, \orgaddress{\street{via Lambruschini 6}, \city{Milano}, \postcode{20156}, \state{Italy}}}

\affil[2]{\orgdiv{Department of Physics ‘Aldo Pontremoli'}, \orgname{Università degli Studi di Milano}, \orgaddress{\street{Via G. Celoria 16}, \city{Milano}, \postcode{20133}, \state{Italy}}}

\affil[3]{\orgdiv{INFN Sezione di Milano and ‘European Theoretical Spectroscopy Facility' (ETSF)}, \orgaddress{\street{Via G. Celoria 16}, \city{Milano}, \postcode{20133}, \state{Italy}}}

%\affil[4]{\orgdiv{Instituto de Física del Litoral}, \orgname{ Consejo Nacional de Investigaciones Científicas y Técnicas, Universidad Nacional del Litoral (IFIS-Litoral, CONICET-UNL)}, \orgaddress{\city{Santa Fe}, \postcode{3000}, \state{Argentina}}}

\affil[5]{\orgdiv{Department of Chemistry}, \orgname{Università di Pavia}, \orgaddress{\street{Via Taramelli 12}, \city{Pavia}, \postcode{27100}, \state{Italy}}}

\affil[6]{\orgdiv{Department of Physics},\orgname{Friedrich-Alexander Universit{\"a}t Erlangen-Nürnberg}, \orgaddress{\street{Erwin-Rommel-Strasse 1}, \city{Erlangen}, \postcode{91058}, \state{Germany}}}

\affil[7]{\orgdiv{Dipartimento di Scienze Chimiche}, \orgname{Università Degli Studi Di Padova}, \orgaddress{\city{Padova}, \postcode{35131}, \state{Italy}}}

\affil[4]{\orgdiv{Instituto de Nanociencia y Materiales de Aragon (INMA)}, \orgname{CSIC-Universidad de Zaragoza}, \orgaddress{\city{Zaragoza}, \postcode{50009}, \state{Spain}}}

\affil[8]{\orgdiv{Departamento de Física de la Materia Condensada}, \orgname{Universidad de Zaragoza}, \orgaddress{\city{Zaragoza}, \postcode{50009}, \state{Spain}}}

\abstract {Graphdiynes are emerging two-dimensional sp-sp$^2$ carbon materials with electronic structures complementing those of graphene, yet their on-surface synthesis is limited by the persistence of metalated intermediates or yields disordered covalent networks. Here, we report an iron-assisted route to covalent hydrogenated graphdiyne monolayers on Au(111) from 1,3,5-tris(bromoethynyl)benzene. Low-temperature scanning tunnelling microscopy, X-ray photoelectron spectroscopy and density functional theory show that Fe scavenges chemisorbed Br byproducts forming FeBr$_2$, in turn promoting the removal of Au adatoms from the organometallic network, thus enabling its metalated-to-covalent conversion under mild thermal treatment. Subsequent annealing removes FeBr$_2$ and yields covalent, ordered domains weakly coupled to the substrate. Scanning tunnelling spectroscopy, combined with density functional theory, reveals a semiconducting gap of about 1.6 eV associated with carbon p$_z$ frontier orbitals. This Fe-mediated on-surface synthesis strategy provides a route to atomically precise, weakly substrate-coupled graphdiyne networks and offers a design principle for two-dimensional carbon semiconductors.} 


\maketitle

\section{Introduction}\label{sec1}
%Graphene sparked interest towards two-dimensional (2D) materials due to its outstanding electronic properties; however, it is a semi-metal, and it lacks a 2D semiconducting counterpart. In this context,
Graphdiynes (GDYs) constitute an emerging class of 2D carbon materials with sp-sp$^2$ hybridization, predicted to show outstanding and tunable electronic and transport properties \cite{baughman1987structure, serafini2021designing} not accessible by graphene.%, complementing graphene. Following the first synthesis by wet chemistry, GDYs have been tested for many advanced applications.\cite{li2019surface,gao2019graphdiyne,li2014graphdiyne,li2010architecture}. However, the connection between atomic-scale structure and electronic properties is limited by the multilayer and defective structure of these systems.

On-surface synthesis (OSS) under ultra-high vacuum (UHV) conditions %emerged as a promising strategy
allows producing single-layer GDYs with atomic precision, also giving access to %insight into
their atomic-scale structure and electronic properties investigation by scanning tunneling microscopy (STM) and atomic force microscopy (AFM). %\cite{sun2016dehalogenative,hu2017recent,bjork2016reaction,li2024theoretical,zhang2012homo,wang2024universal,kang2024surface,wang2024synthesis,fan2015surface,liu2018high,rabia2020structural,d2025unraveling,yang2020metalated,yang2020surface}.
\cite{sun2016dehalogenative,zhang2012homo,wang2024universal,wang2024synthesis,liu2018high,rabia2020structural,d2025unraveling,yang2020metalated,yang2020surface}.
By selecting specifically designed molecular precursors, the temperature-controlled growth of a variety of graphdiyne-like architectures on metal surfaces can be achieved, including 1D nanowires %\cite{lin2017terminal,gao2013glaser,cirera2014synthesis,klappenberger2018functionalized,chen2019hydrogen,li2023pyridinic,zhang2023ullmann,de2024structure,sun2018direct,wang2018kinetic,liu2015lattice,sedona2020surface,cartoceti2025surface}
\cite{klappenberger2018functionalized,zhang2023ullmann,de2024structure,sun2018direct,liu2015lattice,sedona2020surface,cartoceti2025surface}, 2D extended networks %\cite{sun2016dehalogenative,hu2017recent,bjork2016reaction,li2024theoretical,zhang2012homo,wang2024universal,kang2024surface,wang2024synthesis,fan2015surface,liu2018high,rabia2020structural,d2025unraveling,yang2020metalated,yang2020surface,achilli2021graphdiynes} 
\cite{sun2016dehalogenative,zhang2012homo,wang2024universal,kang2024surface,wang2024synthesis,fan2015surface,liu2018high,rabia2020structural,d2025unraveling,yang2020metalated,yang2020surface,achilli2021graphdiynes} and sp-sp$^2$ heterostructures \cite{cartoceti2026graphene}.

%Among the different reactions reported for the OSS of GDYs  \cite{klappenberger2015surface,chen20171d,li2019surface,shang2021surface,xing2023surface,kang2019surface,wu2025reactions,hu2017recent}, 
Ullmann-like dehalogenative homocoupling of halogenated alkynyl precursors %enables the production of atomically precise sp-sp$^2$ carbon nanostructures. This approach 
exploits a noble metal surface as a catalyst to promote the molecules' dehalogenation and their subsequent self-assembly through surface metal adatoms, generating a stable metalated carbon network. Subsequent thermal annealing can remove the adatoms from the structure, promoting the transition to the covalent phase via C-C homocoupling   \cite{clair2019controlling,lackinger2017surface,sun2016dehalogenative,hu2017recent,bjork2013mechanisms,bjork2016reaction,zhou2017surface,li2024theoretical,wang2024universal,cai2010atomically}.
%Despite its promise, the realization of fully covalent and ordered GDYs via Ullmann-like coupling of halogenated alkynyl precursors is limited by metalated intermediates and ubiquitous halogens byproducts. Their removal via thermal annealing is usually accompanied by sp-carbon cross-linking reactions and severe disordering of the network \cite{zhang2012homo,li2024scanning,rabia2019scanning,rabia2020structural,cartoceti2025surface,d2025unraveling}.}
Despite its high-quality fabrication promise, the realisation of fully covalent and ordered GDYs through this method is extremely challenging, since metalated intermediates and reaction byproducts are practically ubiquitous and their removal via thermal annealing is usually accompanied by sp-carbon cross-linking reactions and disordering of the network \cite{zhang2012homo,li2024scanning,rabia2019scanning,rabia2020structural,cartoceti2025surface,d2025unraveling}. In particular, chemisorbed halogen byproducts do not only hinder the network demetalation \cite{cartoceti2026graphene,fan2015surface}, but also affect its electronic properties \cite{d2025unraveling,yang2020metalated}.

Here we show that Fe atoms enable the on-surface synthesis of fully covalent, single-layer, ordered hydrogenated graphdiynes from 1,3,5-tris(bromoethynyl)benzene precursors on Au(111).
Combining low-temperature scanning tunnelling microscopy, X-ray photoelectron spectroscopy and density functional theory, we identify a Fe-mediated Br-scavenging mechanism in which FeBr$_2$ species promote the removal of Au adatoms from the organometallic network, driving its conversion into a covalent framework under mild annealing.
%crucial role of Fe atoms in favouring the removal of Au adatoms from the metalated h-GDY at mild annealing conditions. %Unexpectedly, iron transforms Br atoms from a limiting factor into a co-catalyst of the demetalation process, forming Fe-Br species that enable the controlled removal of metal adatoms at mild annealing conditions, while preserving the structural integrity of the sp-sp$^2$ network. 
%Unlike the metalated counterpart, %we show that the covalent h-GDY is structurally and electronically decoupled from the Au(111) surface. The detailed investigation of its electronic properties shows a band gap of about 1.6 eV defined by C p$_z$ frontier orbitals.%, behaving as freestanding over the Au(111) herringbone reconstruction. 
Notably, the resulting h-GDY is also structurally and electronically decoupled from the substrate, allowing its frontier orbitals (C $p_z$) and semiconducting gap ($\sim$ 1.6 eV) to be resolved.
%Moreover, the almost unperturbed Au Shockley surface electronic state reveals that the covalent h-GDY is also electronically decoupled from the substrate underneath, oppositely to the metalated counterpart.
%This allows for the accurate experimental investigation of its electronic properties, until now limited to theoretical grounds. By combining scanning tunneling spectroscopy (STS) measurements with DFT-calculated local density of states (LDOS), we unveil the electronic character of the covalent h-GDY, resulting close to the freestanding system, with a band gap of about 1.6 eV and dominant orbital contribution of p$_z$ states.
%With this work, we establish an on-surface synthesis route to efficiently obtain covalent, ordered h-GDYs, substantially decoupled from the metal surface %on Au(111), weakly interacting with the substrate and 
%, displaying properties close to the semiconducting freestanding system.
This work establishes Fe-mediated halogen scavenging as a strategy to control demetalation and covalent bond formation in on-surface synthesis of two-dimensional carbon networks.

%These results take a step forward in optimizing the design process of atomically precise 2D all-carbon semiconducting nanostructures. % new perspectives in developing monolayer h-GDYs as a new class of 2D carbon materials with controllable electronic properties synergistic with those of graphene.

%\clearpage
%\section{Results and discussion}\label{sec2}

\section{Results}
\subsection{Fe-mediated on-surface synthesis of covalent h-GDY %promoted by Fe atoms
}
%Although Ullmann-like %on-surface coupling efficiently produces ordered metalated h-GDY frameworks on Au(111), their thermally induced demetalation usually results in the disordering and degradation of the network, preventing the obtainment of fully covalent h-GDYs \cite{rabia2020structural,li2019surface,d2025unraveling,cartoceti2025surface}.%thermal conversion into the fully covalent phase typically leads to disorder and degradation of the acetylenic framework upon removal of the metal adatoms \cite{rabia2020structural,li2019surface,d2025unraveling,cartoceti2025surface}. 
%In the following, we discuss a novel OSS strategy overcoming these limitations.
%Thus, new OSS strategies are required to promote controlled demetalation while preserving the structural integrity of the network.
%\textcolor{red}{In order to achieve a 2D ordered covalent organic network we provide here a protocol that allows us to remove Au adatoms from the metallorganic network without destroying the hexagonal  lattice, as reported in previous works} 

\begin{figure}[h!]
    \centering
   \makebox[\textwidth][c]{%
\includegraphics[width=1.2\textwidth]{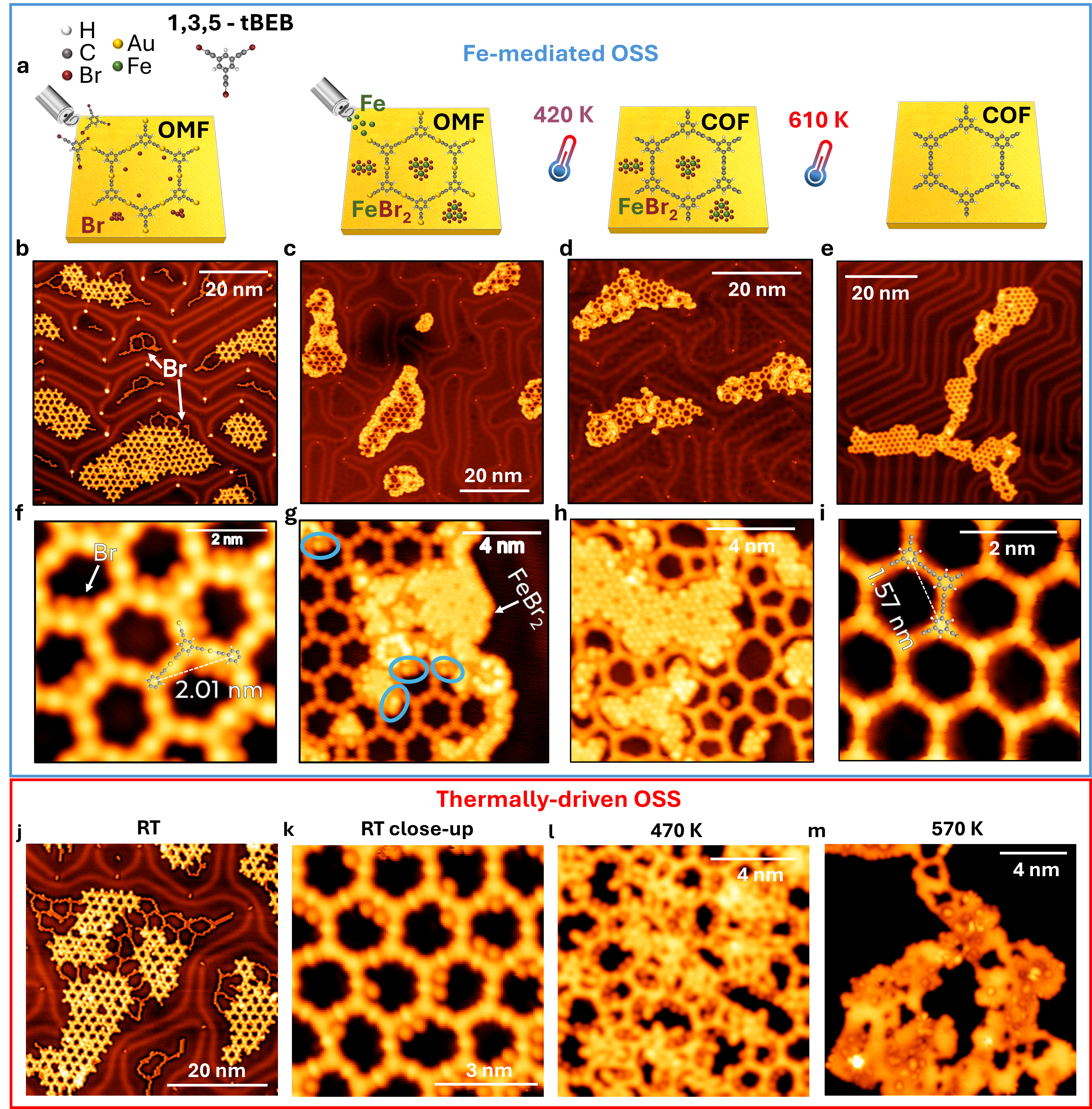}}
    \vspace{1mm}
    \centering
%   \makebox[\textwidth][c]{%
%\includegraphics[width=1.18\textwidth]{Figures/Fig1_synthesis2.png}}
\vspace{5mm}
\captionsetup{width=1.2\textwidth}
    \caption{(a) Sketch of the Fe-mediated OSS process. Above the scheme, the ball-and-stick atomic model of 1,3,5-tris(bromoethynyl)benzene (tBEB) precursor. Schematized steps: tBEB deposition on Au(111) generating the metalated h-GDY; Fe atoms evaporation and bonding with chemisorbed Br atoms resulting in FeBr$_2$ molecules at room temperature; demetalation of the h-GDY; ordering of the covalent h-GDY and FeBr$_2$ disappearance. (b-e) Large-scale and (f-i) close-up atomically-resolved LT-STM images of each reaction step schematized in (a). Ball-and-stick atomic models of the metalated and covalent h-GDY are superimposed on (f) and (i), respectively. Blue circles in (g) represent FeBr$_2$ molecular units. (j-m) LT-STM images of the OSS process upon absence of catalysing Fe atoms, leading to a progressive disruption of the network: (j) Large-scale image of the system at RT, (k-m) Atomically resolved images of the system at RT (k), 470 K (l) and 570 K (m). STM set-point: (b) -500mV/20pA, (c) -80mV/100pA, (d) -2mV/2pA, (e) -1V/5pA, (f) 20mV/180pA, (g) -60mV/100pA, (h) -2mV/10pA, (i) -3mV/30pA, (j) -800mV/50pA, (k) -500mV/50pA, (l) -100mV/100pA, (m) -100mV/250pA.}
    \label{synthesis}
\end{figure}

Deposition of tBEB on Au(111) at room temperature (RT), followed by a mild annealing (400 K), %(refer to Methods for further details). 
form ordered metalated h-GDY islands with a honeycomb-like structure (Figure \ref{synthesis}b,f), where the alternating brighter and fainter spots in LT-STM correspond to gold adatoms and benzene rings
\cite{rabia2019scanning,d2025unraveling,cartoceti2026graphene}.
LT-STM images, their 2D fast Fourier transform (2D FFT) and low-energy electron diffraction (LEED) measurements (Figure S1), reveal a (7$\times$7) superstructure with 0° rotation angle with respect to the underlying Au(111) unit cell, 
%supercell
with a periodicity of 2.01$\pm$0.08 nm (see Figure \ref{synthesis}f) in agreement with the theoretical model \cite{rabia2020structural}.
%These experimental results compare well with the theoretical model, for which a unit cell rotated by 0$^\circ$ degrees has been adopted to minimize the strain (0.5\%) in the calculation cell, commensurate with the Au(111) one to exploit periodic boundary conditions .

%As a by-product of this reaction, 
Br atoms, resulting from tBEB dehalogenation, are found both outside and inside the pores of the metalated h-GDY hexagonal network (Figure \ref{synthesis}b,f). On the surface, they form an extended porous network with local clusters arranged following the $(\sqrt{3}\times\sqrt{3})R30^\circ$ reconstruction (Figure S1) \cite{tao1992situ,zhang2015surface,cartoceti2025surface}.
In the network, they interact with the Au adatoms, stabilizing the system and preventing the network demetalation \cite{cartoceti2026graphene}, as discussed in the following. Therefore, a beneficial effect is expected upon their removal. %when removing Br atoms from the surface. To this end,
%removing these by-products from the surface.
After evaporating a minute quantity ($\sim$0.05 ML) of Fe atoms on the surface, kept at RT, Br atoms are no longer observed in STM, while bright agglomerates form at the edges of the metalated h-GDYs (Figure \ref{synthesis}c,g). %(refer to Methods for further deposition details). 
%Notably, upon Fe evaporation, chemisorbed Br atoms previously present inside and outside the pores of the network %change to 
%can no longer be observed in STM images, while bright agglomerates form at the edges of the metalated h-GDYs %and in the areas uncovered by the network
%(Figure \ref{synthesis}c,g). 
Close-up STM images, complemented by 2D-FFT analysis (Figures S1 h,i), allow us to identify those features as a FeBr$_2$ network, showing a hexagonal lattice with a period of 0.37$\pm$0.02 nm, in reasonable agreement with the value reported in the literature \cite{hadjadj2023epitaxial,xiang2024intrinsically, wilkinson1959neutron}. The Fe-Br chemical bonding due to the formation of FeBr$_2$ is also confirmed by X-ray Photoelectron Spectroscopy (XPS) (Figure S2). %reported in Figure SX, showing the evolution of the binding energy of the Fe$_2p_3/2$ and Br$_3d$ \textcolor{red}{qui metterei qualche detytaglio in più}.
%As exemplified in Figure S1 h, at this stage of the process, FeBr$_2$ occupies the available spaces around the h-GDY, forming a 2D extended network that reproduces the STM structure previously reported in the literature \cite{xiang2024intrinsically}.
%octahedral coordination (1T polytype) displaying the superstructure,
%already observed in the literature
%, due to ordered assembly of Br vacancies

In addition to the extended FeBr$_2$ network, we observe some smaller features in the pores (blue circles in Figure \ref{synthesis}g), which can be identified as single molecular FeBr$_2$ units, as discussed below.

A mild annealing at 420 K for 25 minutes promotes the complete release of Au adatoms from the network, as evidenced by the disappearance of the bright protrusions associated with metal adatoms in LT-STM images of the network (Figure \ref{synthesis}h), and also confirmed by XPS measurements showing the complete disappearance of the C-Au component in the C 1s core level spectrum 
(see Figure S3).
%A mild annealing step at 420 K for 25 min promotes the complete release of Au adatoms from the network, as evidenced by the disappearance of the bright protrusions associated with metal adatoms in LT-STM images (Figure \ref{synthesis}h) and further confirmed by XPS measurements showing the complete disappearance of the C–Au component in the C 1s core-level spectrum (Figure S3).

The transition from the metalated to the covalent phase occurs concomitantly with the partial desorption of FeBr$_2$ (Figure \ref{synthesis}d,h) whose residual phase on the surface is constituted by two coexisting structures, i.e., the FeBr$_2$ 1T polytype with and without vacancies (see Figure \ref{synthesis}h).
%The transition of the system from metalated to covalent is also confirmed by XPS measurements showing the complete disappearance of che C-Au component in the C1s core levels (see Figure S3). 
The presence of two phases, already reported on Au(111) \cite{xiang2024intrinsically}, may indicate an excess of Br atoms or a different local stacking with respect to the underlying substrate. STM images of different regions of the system (Figure S4) unveil the presence of both ordered regions of the carbon network, where the honeycomb-like structure is preserved, and others, where the network is locally deformed into pentagonal, heptagonal and octagonal rings.

A further annealing step at 610 K for 25 minutes completely removes FeBr$_2$ (Figure \ref{synthesis}e,i and Figure S5). Surprisingly, this is accompanied by the overall reordering of the covalent h-GDY (Figure \ref{synthesis}e,i and Figure S5), leading to a hexagonal honeycomb-like network with an average periodicity of 1.57$\pm$0.02 nm (see Figure \ref{synthesis}i and S1). This result is quite unexpected and non-trivial from a chemical standpoint since a large-scale reordering is typically unlikely, as it implies the selective breaking and reformation of C–C bonds.

%Indeed, C–C double bond cleavage of benzene on hematite $\alpha$-Fe\textsubscript2O\textsubscript3 surfaces through strong C-Fe and C-O interactions has been recently predicted \cite{huang2024adsorption}, and previous experimental studies showed the efficient use of Fe-based catalysts to induce the selective oxidative C-C single bond cleavage in alcohols and amines \cite{leonard2021aerobic,mane2021iron,liu2021iron,qi2023water}. The XPS evolution of Fe 2p and Br 3d core levels (Figure S2) shows that, at about 570 K, almost all Br atoms desorb from the surface, meaning that at 570 K FeBr$_2$ is no longer present, while non-negligible metallic Fe traces are still detected. These observations may suggest that these residual Fe atoms can activate the C-C cleavage between the alkynyl units of the covalent h-GDY, which, upon annealing at 610 K, could reform into ordered islands. 
Its mechanistic explanation is beyond the scope of the present work. However, recent works on Fe-induced selective C-C bond cleavage reactions in molecular systems \cite{huang2024adsorption,leonard2021aerobic,mane2021iron,liu2021iron,qi2023water} suggest that transient Fe-mediated bond activation could facilitate network reorganization during the metalated-to-covalent conversion. Moreover, the Co-catalyzed on-surface polymerization of functionalized GDYs has been recently demonstrated, employing coronene as a template for the GDY ordering \cite{shu2026synthesis}. %demonstrates the cobalt-catalyzed on-surface 2D covalent polymerization of fluorographdiyne nanosheets, and theoretical studies have predicted C–C bond activation of aromatic molecules on iron-oxide surfaces through strong Fe–C and Fe–O interactions \cite{huang2024adsorption}, while Fe-based catalysts are known to promote selective C–C bond cleavage reactions in molecular systems \cite{leonard2021aerobic,mane2021iron,liu2021iron,qi2023water}. These observations 
%From XPS we know that %The XPS evolution of Fe 2p and Br 3d core levels (Figure S2) shows that, 
In our system, at about 570 K, almost all Br atoms desorb from the surface, %indicating that FeBr$_2$ is no longer present, 
while non-negligible metallic Fe traces are still detected (Figure S2). These residual Fe species %, in the absence of Br,
may %influence the stability of the covalent network and potentially 
contribute to C–C bond reorganization during annealing at 610 K, leading to the formation of ordered islands without the aid of any templating structure.
%To demonstrate the crucial improvement established by our Fe-catalyzed OSS with respect to the typical thermally activated Ullmann-like coupling, 
 
 %Figure \ref{synthesis}j-m shows the same 1,3,5-tBEB-based h-GDY network obtained on Au(111) through the typical thermally activated OSS, without the use of iron. The first step consists of the obtainment of a metalated, ordered h-GDY at room temperature upon molecules' sublimation on gold (Figure \ref{synthesis}j). However, without Fe, annealing at 470 K (Figure \ref{synthesis}k,l) results in the disruption of the network, with most of it still metalated and all Br atoms chemisorbed on the surface. Proceeding further with annealing at 570 K (Figure \ref{synthesis}m) results in the conversion of the h-GDY to a completely amorphous system, where the original honeycomb-like structure can no longer be recognized. Conversely, in the presence of Fe atoms, annealing at 420 K (Figure \ref{synthesis}d,h and Figure S4) results in the complete conversion of the network from metalated to covalent while preserving its structural integrity, and the subsequent annealing at 610 K is even beneficial for the overall reordering of the covalent network (Figure \ref{synthesis}e,i and Figure S5).
Compared to the typical thermally activated Ullmann-like coupling (Figure \ref{synthesis}j-m), our Fe-mediated OSS approach (Figure \ref{synthesis}a-i) enables the complete network demetalation at a lower temperature, also ensuring a higher thermal stability of the fully covalent system. Without Fe, annealing at 470 K (Figure \ref{synthesis}l) already results in the disruption of the network before achieving its complete demetalation, and further annealing at 570 K (Figure \ref{synthesis}m) results in a completely amorphous carbon system, where the original structure can no longer be recognized.

\clearpage
\newgeometry{
    top=0.8cm,
    bottom=4cm,
    right=3cm
}

\restoregeometry

\clearpage
\subsection{Covalent conversion mechanism via Fe-mediated Br scavenging}
We here unravel the mechanism underlying the Fe-mediated metalated-to-covalent transition, correlating the evolution of the carbon network to that of Br-derived structures identified on the surface during the annealing.
%, discussing their role in the removal of the Au adatoms from the metalated h-GDY.

\begin{figure}[h!]
\centering
\includegraphics[width=1.0\textwidth]{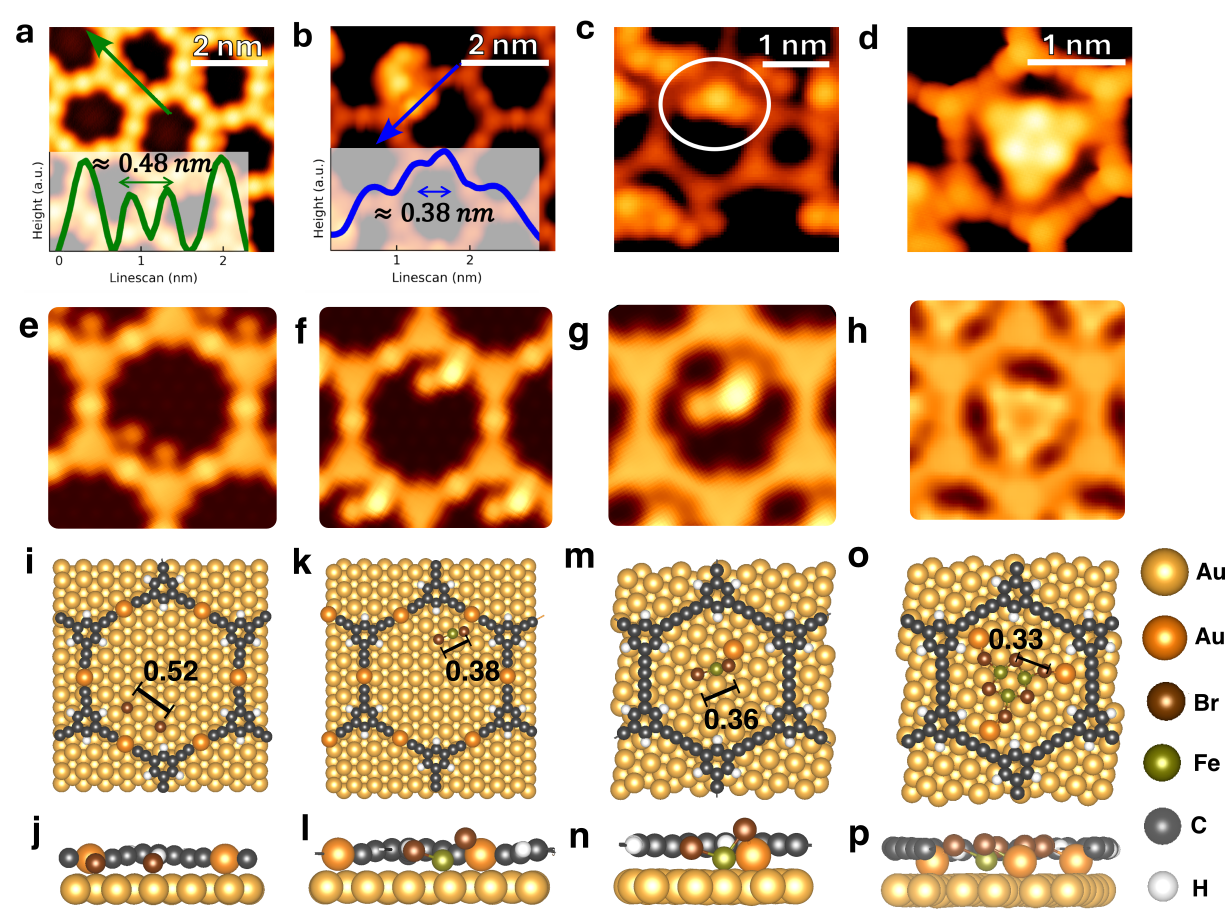}
\caption{Atomically-resolved LT-STM image of: (a) metalated h-GDY before Fe atoms evaporation. Green dashed arrow identifies a pair of chemisorbed Br atoms arranged according to the $\sqrt{3}\times \sqrt{3}$ reconstruction, whose height profile is reported as inset below; (b) h-GDY after Fe evaporation and mild annealing at 385 K. Blue dashed arrow identifies a bright-faint protrusion pair, whose height profile is reported as inset below; distances have been extracted from line profile measurements performed over different high-resolution LT-STM images of the metalated h-GDY. (c) a demetalated hexagon of the h-GDY after annealing at 385 K. White circle exemplifies the Au-stabilized FeBr$_2$ unit. (d) FeBr$_2$ nanocluster confined into a pore of the covalent h-GDY. (e-h) STM simulations  and corresponding relaxed structural models in top view (i,k,m,o) and side view (j,l,n,p) of the experimental images shown in the first row (a-d). STM setpoint: (a)20mV/180pA, (b)-2V/160pA, (c)-2V/160pA, (d) -2.5mV/30pA. STM constant current simulations: integration range [E$_F$-0.5,E$_F$], isosurface $10^{-6}$ states/a$_0^3$.
}\label{FeBr}
\end{figure}

The tBEB dehalogenation involves the release of three Br atoms per molecule, leading to a total of six Br atoms per unit cell containing two molecules. At low coverage, an average of 2.3 Br atoms per pore is observed, preferentially located near Au adatoms within the network (see Figure \ref{FeBr}a). This is confirmed by our theoretical calculations, indicating that a single Br preferentially adsorbs at an FCC site of the underlying Au(111) surface, and is stabilized by proximity to a gold adatom (see Figure S7)%, as the one at the edge of the h-GDY
. Moreover, Br adatoms tend to distribute over different pores (see Figure S6), although in some of them they cluster into an incipient $\sqrt{3}\times \sqrt{3}$ reconstruction%, already observed for Br on Au(111) 
(Figure S1e, cyan rhombus, and corresponding model in S6a).

At this stage, Br pairs inside the metalated h-GDY pores show a uniform contrast, with experimental interatomic distance (0.48$\pm$0.02 nm, Figure \ref{FeBr}a) fitting the lateral spacing between second nearest neighbours FCC adsorption sites of gold.
This value is well reproduced by the structural model and simulated STM images (Figure \ref{FeBr}e,i), with a small discrepancy arising from the calculated Au lattice constant (4.21 \AA) relative to the experimental value (4.08 \AA). Notably, the metalated h-GDY is corrugated due to the relevant interaction between Au adatoms and the surface (Figure \ref{synthesis}j).

Following Fe deposition and annealing at 385 K, the %protrusions previously assigned to 
paired Br show an asymmetric contrast%, particularly at the pore edges 
and a reduced interatomic distance ($0.38\pm0.01$ nm, Figure \ref{FeBr}b)%Concomitantly, their average separation decreases to $0.38\pm0.01$ nm (inset of Figure \ref{FeBr}b)
, indicating that the Br atoms no longer follow the Au(111) registry.

%The experimental features are 
Simulations show that both observations are consistent with the formation of a FeBr$_2$ molecular unit stabilized by interaction with an Au adatom at the edge of the pores (Figure \ref{FeBr}k,l and S8, S9).
%, where one Br atom interacts with an Au adatom. 
%The simulated STM image (Figure \ref{FeBr}f) reproduces both the asymmetric contrast observed in Figure \ref{FeBr}b and the reduced Br-Br separation relative to Figure \ref{FeBr}a (model in Figure \ref{FeBr}e).

%The structural model that best reproduces these experimental observations consists of an iron atom coordinated to two Br atoms (Figure S8, S9), one of which interacts with an Au adatom at the edge of the metalated h-GDY pore (Figure \ref{FeBr}k,l). The correspondence between the simulated STM image for FeBr$_2$ near the pore edge (Figure \ref{FeBr}f) and the experiment (Figure \ref{FeBr}b) supports our interpretation, as well as their separation reduction compared to Figure \ref{FeBr}a (whose model is reported in Figure \ref{FeBr}e).
In particular, the contrast asymmetry originates from the different adsorption heights of the two Br atoms. They are both bound to a deeper Fe adatom located at 2\AA~from the surface (see Figure \ref{FeBr}l); one Br interacting with the substrate Au atoms and located at 2.9 \AA~ from the surface, and the other displaced upward to 3.5 \AA~upon FeBr$_2$ formation, interacting with the Au adatom embedded in the metalated h-GDY (Figure \ref{FeBr}l).

This interaction stabilizes the pore-edge configuration (Figure S10a,b) and promotes the Au stripping from the metalated network with an energy gain of 0.8 eV (Figure S10c,d), being the reason for the overall conversion of the h-GDY to the covalent phase.

%Upon demetalation, the Au adatom is expected to maintain the interaction with the FeBr$_2$ unit within the pore, with relative atomic heights and distances preserved, as confirmed by the relaxed structure in Figure \ref{FeBr}m. The simulated STM image resulting from this model (Figure \ref{FeBr}g) shows an elongated structure with variable brightness, observable also in STM images, exemplified by Figure \ref{FeBr}c. The three imaged spots are the innermost Br atom, the outermost Br atom and the Au adatom, blurred, which has been removed from the h-GDY. The lateral view of the model (Figure \ref{FeBr}n) shows the same atomic arrangement as Figure \ref{FeBr}l, with a remarkable flattening of the h-GDY network, now resembling a rather detached freestanding system (corrugation passing from 0.32 \AA~ to less than 0.05 \AA).

Upon demetalation, the Au adatom remains coordinated with the FeBr$_2$ unit within the pore (Figure \ref{FeBr}m). 
The simulated STM image (Figure \ref{FeBr}g) reproduces the elongated feature observed experimentally (Figure \ref{FeBr}c), where the three spots correspond to the two Br atoms and the stripped Au adatom.%, the latter appearing as a blurred protrusion.
%Concomitantly, the all-carbon h-GDY network undergoes an almost complete flattening consistent with the loss of Au adatoms, with the corrugation decreasing from 0.32 \AA~to below 0.05 \AA~(Figure \ref{FeBr}n).

%Additional argument supporting this interpretation is provided by the more complex triangular features observed by STM within the pores, as the one reported in Figure \ref{FeBr}d. We assign this structure to a cluster of three FeBr$_2+$Au units, resulting in the triangular complex reported in Figure \ref{FeBr}o. This structure resembles the one proposed for low coverage CoBr$_2$ \cite{CoBr2}, where the role of Au adatoms in stabilizing the complex was recognized, also contributing to the observed STM pattern. The agreement between simulated (Figure \ref{FeBr}h) and experimental (Figure \ref{FeBr}d) STM images supports this assignment, confirming that Au adatoms are located near the triangular structure, in proximity to the h-GDY edges, but removed from the h-GDY network.

Further support for this assignment is provided by the triangular features occasionally observed within the pores (Figure \ref{FeBr}d), well reproduced by simulated STM (Figure \ref{FeBr}h). These structures are consistent with a cluster of three FeBr$_2$+Au units arranged as shown in Figure \ref{FeBr}o. Similar complexes have been proposed for low-coverage CoBr$_2$ on Au surfaces \cite{CoBr2}, where Au adatoms were found to stabilize the assembly and contribute to the STM pattern. %The agreement between simulated (Figure \ref{FeBr}h) and experimental (Figure \ref{FeBr}d) STM images further supports this assignment, indicating that the Au adatoms remain near the h-GDY pore edges while being detached from the network.

%The combined STM and DFT analysis of the process dynamics taking place during the Fe-catalysed OSS unveils not only the fundamental role of Fe atoms but also the dual role played by Br atoms in allowing the removal of gold adatoms from the network. While in the absence of Fe, chemisorbed Br atoms are stabilized in the proximity of Au adatoms of the network and hinder the transition of the h-GDY from metalated to covalent \cite{cartoceti2026graphene,d2025unraveling,rabia2020structural}, in the presence of Fe atoms, Br atoms form FeBr$_2$ and bind to Au adatoms that stabilize the structure, stripping them from the network and, thus, enabling its transition to the covalent phase.
The combined STM and DFT analysis highlights the dual role of Br in the Fe-mediated OSS process. In the absence of Fe, chemisorbed Br atoms stabilize Au adatoms and hinder the conversion %of h-GDY from the metalated 
to the covalent phase \cite{cartoceti2026graphene,d2025unraveling,rabia2020structural}. Upon Fe incorporation, Br form FeBr$_2$ species that coordinate Au adatoms, stripping them from the network, thereby promoting the formation of fully covalent h-GDY.
\clearpage

\clearpage
\subsection{Structural and electronic properties of the covalent h-GDY network}
%The quality of the covalent h-GDY obtained via our Fe-catalyzed OSS process, being ordered and free from reaction byproducts, allowed us to accurately investigate its structural and electronic properties, and to compare them with the metalated counterpart.
%The conversion from metalated to covalent h-GDY weakens the coupling with the substrate, which is mediated by the interaction between Au adatoms and the surface layer. As a consequence, remarkable differences in the properties of covalent and metalated h-GDY can be observed.

The ordered and clean nature of the covalent h-GDY produced by Fe-mediated OSS enables a detailed characterization of its structural and electronic properties. The conversion to the covalent phase weakens the Au-adatom-mediated coupling to the substrate, resulting in marked differences from the metalated network.

Upon conversion from metalated to covalent phase, LT-STM images reveal a reduction of the unit-cell periodicity (Figure \ref{synthesis}f,i), described by a simulated unit cell rotated by $8.9^\circ$, requiring only 0.7\% strain to achieve commensurability with Au(111) under periodic boundary conditions \cite{rabia2020structural}. Consistently with previous calculations \cite{rabia2020structural}, Au-adatom removal weakens the substrate interaction, resulting in a flatter network and a larger separation from the surface (Figure \ref{FeBr}n,p).

Further evidence of the progressive decoupling from the substrate is provided by the recovery of the Au(111) herringbone reconstruction, which is largely lifted by the formation of the metalated h-GDY (Figure \ref{synthesis}b) 
%owing to both the $(\sqrt{3}\times \sqrt{3})$R$30^\circ$ Br overlayer and the Au-adatom-mediated interaction between the network and the substrate
\cite{basagni2015molecules,cartoceti2025surface}. 
The herringbone remains lifted after Fe deposition (Figure \ref{synthesis}c) and Au-adatom removal at 420 K (Figure \ref{synthesis}d), due to the presence of FeBr$_2$. Complete recovery is observed only after annealing at 610 K (Figure \ref{synthesis}e), upon removal of FeBr$_2$, providing direct evidence of the weak coupling between covalent h-GDY and Au(111).

The average lateral size of the islands increases from $\sim$24 nm at 420 K to $\sim$43 nm at 610 K (Figure S4, S5), accompanied by the emergence of branched structures. We attribute this behaviour to the enhanced mobility of the decoupled covalent domains, which 
%once freed from both Au adatoms and FeBr$_2$, 
can merge into larger domains. 
%upon thermal annealing.
%\textcolor{red}{(statistics performed over several domains exemplified in Figure S4, S5 shows an average lateral extension of $\sim$24 nm at 420 K and $\sim$43 nm at 610 K)}

%Another crucial indication of a substantially different interaction with the substrate between metalated and covalent h-GDY is encoded in the energy position of the Au(111) surface state (SS), located at about -0.45 eV below the Fermi level on the pristine surface \cite{andreev2004adsorbed}. Our STS measurements on metalated h-GDY show a relevant energy shift for the Au(111) surface state when confined in the hexagonal pores of the network with respect to the pristine state, up to +0.4 eV on Au(111) (see Figure S11-S12). This is in agreement with previous reports on metalated h-GDY on Au(111) and Ag(111)  \cite{li2024scanning,d2025unraveling,yang2020metalated}. In contrast, upon complete conversion to covalent h-GDY, the Au(111) surface state confined in the hexagonal pores appears at -0.35 eV (see grey curve in Figure \ref{STS-PDOS}e and Figure S11) with a limited shift with respect to its pristine Au(111) value. This can be more easily visualized in Figure S11, where the direct comparison between the two cases is shown.

\begin{figure}[h!]
\centering
\makebox[\textwidth][c]{%
\includegraphics[width=1\textwidth]{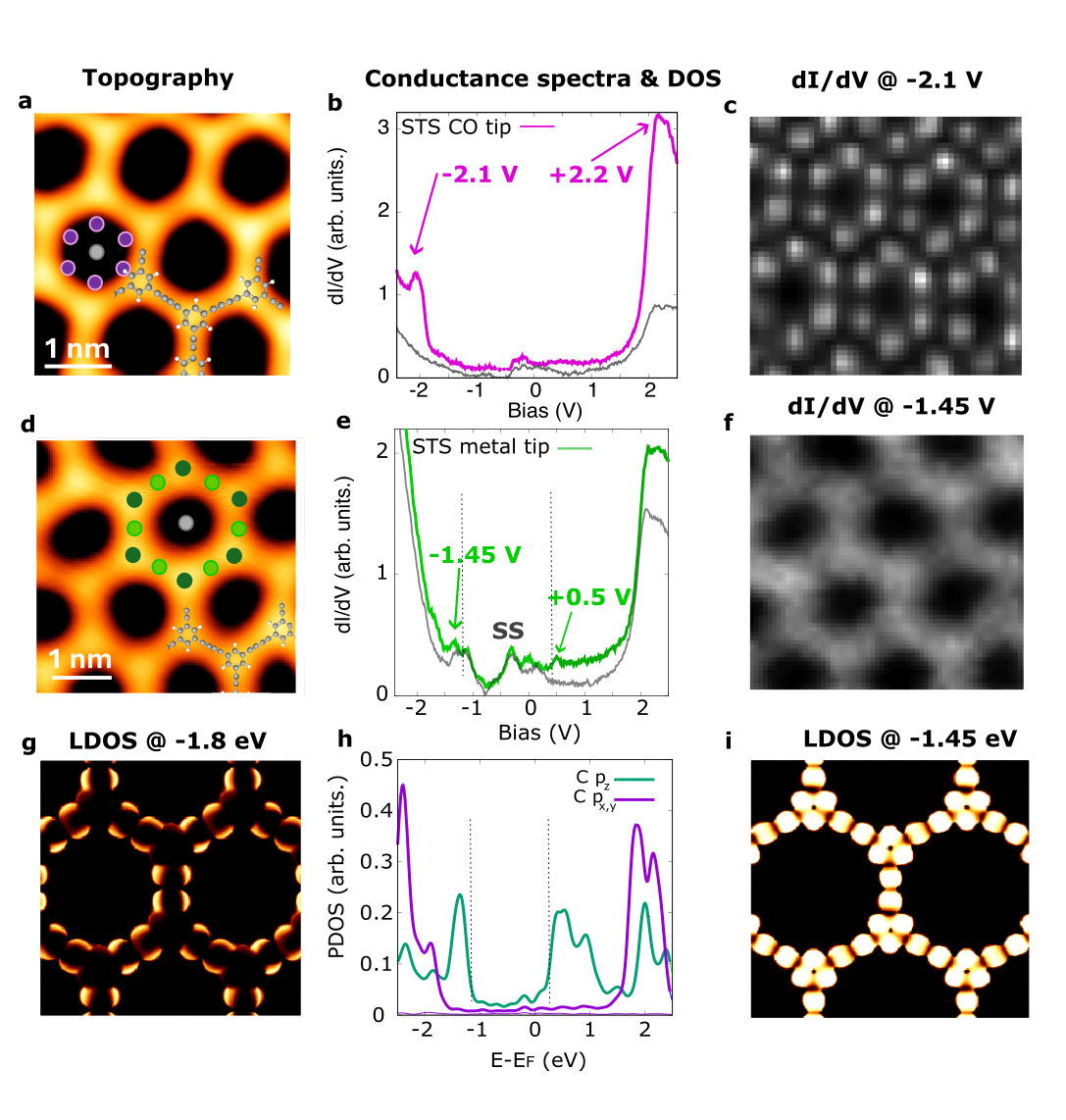}}
%\captionsetup{width=1.2\textwidth}
\caption{(a) Atomically resolved LT-STM image of the STS grid area acquired with CO-functionalized tip. (b) Conductance spectra acquired at the colour-coded dot positions in (a): grey, central pore position; violet, average spectrum marked by the violet dots in (a), with two prominent spectral features at -2.1 V and +2.2 V.
(c) Constant-current dI/dV map acquired with CO-functionalized tip at -2.1 V. STS/STM set-point for (a-c): -0.55 V/40 pA, with lock-in amplitude 20 mV and frequency 817 Hz. (d) Atomically resolved LT-STM image of the STS grid area acquired with metallic tip. (e) Conductance spectra acquired at the colour-coded dot positions of (d): grey, central pore position; green, average of the STS point spectra acquired on the carbon network, specifically on the marked phenyl rings (dark green dots) and acetylenic links (light green dots), both characterized by two prominent spectral features at -1.45 V and +0.5 V. (f) Constant-current dI/dV map acquired with metallic tip at -1.45 V. STM/STS set-point: -0.55 V/50 pA, with lock-in amplitude 20 mV and frequency 817 Hz. (g) and (i) simulated dI/dV maps at -1.8 eV and -1.45 eV, respectively, considering the height modulation introduced by the feedback setpoint. (h) Orbital projected density of states (PDOS) of carbon atoms in h-GDY.
}\label{STS-PDOS}
\end{figure}

An additional crucial indication of the different substrate-coupling in metalated and covalent h-GDYs is the Au(111) surface state (SS). On pristine gold, it is located at $\sim$-0.45 eV \cite{andreev2004adsorbed}; in the metalated h-GDY it shifts up to +0.4 eV, also in agreement with previous reports \cite{li2024scanning,d2025unraveling,yang2020metalated} (Figure S11,S12), while in the covalent h-GDY it is observed much closer to its nominal energy (i.e. -0.35 eV, Figure \ref{STS-PDOS}e and Figure S11), indicating a substantially weaker electronic interaction with the substrate. %.In metalated h-GDY, the SS confined within the hexagonal pores is shifted by up to +0.4 eV relative to Au(111), consistent with previous reports \cite{li2024scanning,d2025unraveling,yang2020metalated} (Figure S11,S12). In contrast, after complete conversion to covalent h-GDY, the confined SS is observed at -0.35 eV (grey curve in Figure \ref{STS-PDOS}e and Figure S11), i.e. much closer to its pristine Au(111) energy, indicating a substantially weaker interaction with the substrate.

%The recovery of the herringbone reconstruction and the changes observed in the Au(111) surface electronic properties at the Fermi level confirm that the conversion of the h-GDY from the metalated to the covalent phase is accompanied by a substantial decoupling of the network from the substrate, which is also expected to affect its overall electronic properties, as discussed in the following.  
%The recovery of the herringbone reconstruction together with the evolution of the Au(111) surface state confirms that the transition from metalated to covalent h-GDY is accompanied by substantial substrate decoupling, which in turn profoundly modifies the electronic properties of the network, as discussed below.

%Oppositely to its metalated counterpart, 
This substantial decoupling of the covalent h-GDY from Au(111) allows us to precisely determine its electronic properties, closely matching the simulated freestanding system.

%To unveil the electronic properties of the covalent h-GDY, we performed conductance (dI/dV) spectroscopy at 4.8 K by measuring constant current spectroscopic grids in the -2.5 to +2.5 V energy range over the two 5 x 5 nm${^2}$ sample areas of the covalent network shown in the LT-STM image of Figure \ref{STS-PDOS}a,d. The STS curves reported in Figure \ref{STS-PDOS} were acquired with a CO-functionalized tip on the cove positions of the network, and with a metallic tip on top of the network, while the complementary measurements are reported in Figure S13. The use of two different tip terminations allows for the enhancement of different orbital features: indeed, the metallic tip states are of mainly s-wave character, resulting in an enhancement of $p_z$ related features, while the addition of CO molecule introduces also tip states having $p_x$- and $p_y$-wave character \cite{paschke2025distance} (see Figure S13 and relative discussion).
%To probe the electronic properties of covalent h-GDY, we 

We performed conductance dI/dV spectroscopy at 4.8 K by acquiring constant-current grids over the two 5$\times$5 nm$^2$ regions shown in Figure \ref{STS-PDOS}a,d, within the energy range from $-$2.5 to $+$2.5 V. The spectra reported in Figure \ref{STS-PDOS} were acquired at the cove sites using a CO-functionalized tip and on top of the network using a metallic tip (complementary measurements are shown in Figure S13). The two tip terminations enhance different orbital contributions: metallic tips emphasize p$_z$-derived features, whereas a CO-functionalized tip introduces also p$_{x,y}$-wave components \cite{paschke2025distance} (see Figure S13).

%The point spectra acquired on the cove positions of the networks with a CO-functionalized tip (violet dots in Figure \ref{STS-PDOS}a), combined with differential conductivity maps acquired in constant current mode, revealed two relevant features at -2.1 V and +2.2 V (violet line in Figures \ref{STS-PDOS}b and complementary data with metallic tip in Figure S13 e).
%The differential conductivity map at -2.1 V, which matches with the one obtained at +2.2 V (see also Figure S14 a), is shown in Figure \ref{STS-PDOS}c, 
Point spectra acquired at the cove sites with a CO-functionalized tip (violet dots in Figure \ref{STS-PDOS}a), together with dI/dV maps acquired in constant current mode, reveal two prominent features at -2.1 V and +2.2 V (violet curve in Figure \ref{STS-PDOS}b; see also Figure S13e). The corresponding conductance maps are nearly identical (Figure \ref{STS-PDOS}c and Figure S14a) and reveal maximum local density of states (LDOS) in correspondence of the H-terminated vertices of the phenyl rings.

Point spectra acquired on the h-GDY backbone with a metallic tip (green dots in Figure \ref{STS-PDOS}d), together with constant-current dI/dV maps, reveal additional features at -1.45 V (with a shoulder at -1.6 V) and +0.5 V (green curve in Figure \ref{STS-PDOS}e; see also Figure S13b). These features become more evident when compared to spectra acquired at the pore centre (grey curve in Figure \ref{STS-PDOS}e). The corresponding differential conductivity maps at -1.45 V and +0.5 V are nearly identical (Figure \ref{STS-PDOS}f and Figure S14b) and display maximum LDOS on the h-GDY backbone.
The relative intensity of the spectral features depends on the setpoint used during grid acquisition, which determines the tip-sample distance. As discussed in the Supporting Information (``Insight into LT-STS measurements''), the tip approaches more closely the H-terminated sites than the carbon backbone, thereby modulating the conductance contrast.

%On the other hand, the point spectra acquired over the covalent h-GDY network by using a metallic tip (green dots in Figure \ref{STS-PDOS}d), combined with differential conductivity maps acquired in constant current mode, evidence additional contributions. In particular, a small peak at -1.45 V, ending up in a shoulder at -1.6 V, and another peak at +0.5 V (green line in Figures \ref{STS-PDOS}e and complementary data with CO-tip in Figure S13 b), more evidently emerging from a comparison with the averaged point spectrum acquired at the center of the h-GDY pore (gray line in Figure \ref{STS-PDOS}e).
%The differential conductivity map at -1.45 V, which resembles the one at +0.5 V (reported in Figure S14 b), is reported in Figure \ref{STS-PDOS}f, and reveals maximum LDOS in correspondence of the h-GDY network.
%It is worth noticing that the conductance spectra are sensitive to the setpoint that defines the height modulation of the tip during the grid acquisition. In this way,tip–sample distance is smaller above H atoms and larger above the backbone C atoms  (see ``Insight into LT-STS measurements" section, Supporting Information). 

To assign the experimentally observed STS features to specific atomic orbitals, we compare STS spectra with the orbital-projected density of states of covalent h-GDY (Figure \ref{STS-PDOS}e).
The experimental features at -1.45 eV and 0.5 eV match well with the calculated p$_z$ states in the PDOS (green line in Figure \ref{STS-PDOS}h), whereas the experimental peaks at -2.1 eV and +2.2 eV (Figure \ref{STS-PDOS}b)
are assigned to (p$_{x,y}$) states, after a minor rescaling of the unoccupied states to account for the known limitations of DFT.
%can be identified as p$_{x,y}$ states from the comparison with Figure \ref{STS-PDOS}h, upon a mild energy rescaling of empty states, compatible with the typical failure of DFT.

This assignment is further supported by calculated LDOS maps at -1.8 eV (Figure \ref{STS-PDOS}g) and -1.45 eV (Figure \ref{STS-PDOS}i), obtained at fixed bias and constant tip-sample distance (see Methods). In agreement with the experimental dI/dV maps, the -1.8 eV state is suppressed on the network, whereas the -1.45 eV state exhibits enhanced intensity on the carbon backbone, confirming its (p$_z$) character.

%The analysis of experimental dI/dV maps is also supported by the calculated local density of states (LDOS) maps at -2.1 eV (Figure \ref{STS-PDOS}g) and -1.45 eV (Figure \ref{STS-PDOS}i), computed by sampling the LDOS at fixed bias and constant distance on a z(x,y) profile (see Methods).
%The LDOS maps confirm that states at -2.1 eV are dark on the network while states at -1.45 eV show a bright signal on the carbon atoms of the h-GDY, in agreement with the experiment, and confirming the assignment to out-of-plane orbitals.

%Based on this analysis, the overall electronic character of covalent h-GDY can be classified as semiconducting, despite the small contributions of states around the Fermi level that are due to a residual mild interaction between carbon $p_z$  orbitals and $s,p$ states of the surface.
%The experimental HOMO-LUMO gap of the covalent network on Au can be set to $\sim 1.6$ eV by considering as gap edges the rising of the features marked in Figure \ref{STS-PDOS}e (dashed lines). This value is in good agreement with the value extracted from the PDOS, i.e. $\sim 1.2$ eV (dashed lines in Figure \ref{STS-PDOS}h) in the limit of the failure of DFT. Supporting calculations with hybrid functionals provide a still better agreement with the experiment (HOMO-LUMO gap $\sim 1.45$ eV, see Figure S15).
%Our estimated gap is smaller than the one reported in the literature \cite{li2024scanning} for the same carbon network on Au(111). Indeed, in that case, the frontier orbitals were assigned to the features at $\sim\pm 2$ V, disregarding the $p_z$ states near the Fermi level.

Overall, covalent h-GDY exhibits a semiconducting character, despite a small residual density of states around the Fermi level arising from weak hybridization between carbon p$_z$ orbitals and Au s,p states. The experimental HOMO-LUMO gap, estimated from the onset of the frontier-state features in Figure \ref{STS-PDOS}e (dashed lines), is $\sim$1.6 eV. This value is consistent with the gap obtained from the calculated PDOS ($\sim$1.2 eV; Figure \ref{STS-PDOS}h), as well as with hybrid-functional calculations ($\sim$1.45 eV; Figure S15).
Our estimated gap is smaller than the value previously reported for the same network on Au(111) \cite{li2024scanning}. We attribute this difference to the assignment of the frontier orbitals: 
%whereas Ref.~\cite{li2024scanning} associates them with features at $\sim\pm$2 V, 
our combined STS and DFT analysis identifies the p$_z$-derived states closer to the Fermi level as the relevant HOMO and LUMO states.

%Comparison with the metalated network (Figure S12c,d) highlights the electronic consequences of the metalated-to-covalent transition. In the covalent phase, the density of states at the Fermi level is strongly reduced, consistent with a more pronounced semiconducting character, while the gap of the in-plane (p$_{x,y}$) states is fully restored. The transition is also accompanied by a weaker hybridization between the h-GDY p$_z$ orbitals and the substrate, despite an overall charge-transfer-induced energy shift \cite{rabia2020structural}. Consistently, both bias-dependent dI/dV maps and PDOS calculations reveal a shift of carbon-derived states towards higher energies in the covalent network (Figure S12a,c,d).

Comparison between STS of covalent and metalated network (Figure S12a,b) and related PDOS  (Figure S12d,e) evidences an overall reduction of the states at the Fermi level upon metalated-to-covalent transition. The covalent phase shows a more pronounced semiconducting character, the gap of the in-plane (p$_{x,y}$) states is restored and the hybridization between the h-GDY p$_z$ orbitals and the substrate is reduced resembling the freestanding case (Figure S12a). Due to an overall charge transfer \cite{rabia2020structural} both bias-dependent dI/dV maps and PDOS calculations reveal a shift of carbon-derived states towards lower energies, more marked in the covalent network. 

%From a comparison with our calculation on the metalated graphdyine (Figure S12d), we notice a reduction of the density of state at the Fermi level in the purely covalent system, suggesting a more remarkable semiconducting character, compared to the metallic one of the metalated phase. In the covalent network the gap of in plane $p_{x,y}$ orbitals is completely restored (Figure S12 c,d). The transition from the metalated to the covalent structure is also accompanied by a reduction of the hybridization between the carbon $p_z$ orbitals of the h-GDY network and the substrate (see also ref. \cite{rabia2020structural}) even though they are shifted in the latter by an overall charge transfer (Figure S12 c,d). Moreover, the analysis of bias-dependent dI/dV maps of the metalated network in comparison with the one performed for the covalent network reveals an overall shift of carbon-related features toward higher energies (see Figure S12a) which is confirmed by the PDOS analysis (see Figure S12 c,d). 

The differences between covalent and metalated h-GDY demonstrate that the Fe-driven, Br-mediated OSS process provides effective control over the structural and electronic properties of the network. In addition, the resulting all-carbon h-GDY remains stable up to 610 K, substantially exceeding the thermal stability of the metalated phase and enabling improved structural ordering.

The combination of the network order, weak substrate coupling, and finite band gap establishes covalent 2D h-GDYs as a promising platform for future all-carbon electronic and optoelectronic devices.

%The reported differences between covalent and metalated h-GDY evidence the possibility to achieve a material with substantially different properties by exploiting our Fe-driven and Br-mediated OSS process. Furthermore, the obtained all-carbon h-GDY reveals thermal stability up to 610 K, which is substantially higher than its metalated counterpart, with a benefit for its reordering.  
%The combination of structural order, substrate decoupling, and a finite band gap makes this material a promising candidate for future all-carbon electronic and optoelectronic applications.

\newpage
\thispagestyle{empty}

\clearpage
\section{Conclusions}
This work establishes a Fe-mediated on-surface synthesis strategy for obtaining covalent, single-layer hydrogenated graphdiyne on Au(111). By combining low-temperature scanning tunnelling microscopy, X-ray photoelectron spectroscopy and density functional theory, we identify a Br-scavenging mechanism in which Fe forms FeBr$_2$ species that promote the removal of Au adatoms from the organometallic network, enabling its conversion into a covalent framework under mild annealing. Removal of FeBr$_2$ upon further annealing restores structural order and yields h-GDY domains that are weakly coupled to the Au(111) substrate, allowing the frontier carbon p$_z$ states to be resolved, revealing a semiconducting gap of about 1.6 eV. 
%This is of crucial importance to assess its electronic properties, which are only slightly perturbed with respect to the freestanding case, bearing a semiconducting character with a bandgap of about 1.6 eV. %Indeed, the bandgap is defined by p$_z$ states delocalized over the carbon covalent structure. 
More broadly, these results show that controlled halogen scavenging can direct demetalation and covalent bond formation in on-surface synthesis, providing a design principle for atomically precise two-dimensional carbon semiconductors.
%This study opens new pathways for the design of high-quality, atomically thin semiconducting sp-sp$^2$ covalent networks that could complement graphene in a wider class of 2D carbon materials with structurally engineered electronic properties having huge potential for future semimetallic-semiconducting all-carbon nanojunctions.

\clearpage
\section{Methods}\label{sec11}

\subsection*{Experiment}
All the experiments are carried out under ultra-high vacuum conditions, with a base pressure below $10^{-10}$~mbar. Single-crystal Au(111) samples (MaTeck GmbH and SPL) are used for the on-surface synthesis, following several cleaning cycles of Ar$^+$ sputtering and annealing at 770~K. 1,3,5-tris(bromoethynyl)benzene (tBEB) molecular precursor is loaded in powder form in a Knudsen cell and evaporated for 15-20 seconds over Au(111), keeping the crucible at 300 K and the substrate at room temperature, reaching a chamber pressure of $\sim$2.5~$\times~10^{-10}$~mbar. Upon surface-catalysed halogen cleavage, tBEB molecules undergo coupling reaction through the incorporation of gold adatoms from the substrate, generating a metalated h-GDY \cite{d2025unraveling, sun2016dehalogenative, yang2020metalated, shu2020atomic, cartoceti2025surface}. To improve the order of the metalated h-GDY, a mild annealing at about 400 K is performed.
Fe atoms are evaporated on the obtained system through an e-beam evaporator equipped with an integral flux monitor. The sample is maintained at room temperature while Fe atoms are deposited on top, setting the emission current I$_{em}$ to 13.7 mA, the ion flux to 0.1 nA and the deposition time to 80 seconds. A constant pressure of 2.5$\times10^{-10}$~mbar during Fe evaporation ensures the perfect cleanliness of the deposit, without contamination from the metallic rod. After Fe deposition, a sequential annealing at 420 K and 610 K results in the generation of a completely covalent two-dimensional h-GDY network. %STM measurements were carried out with two Omicron microscopes, one operating at room temperature (RT) and the other at 4.8 K (low temperature, LT). 
Low temperature (LT) scanning tunneling microscopy and spectroscopy (LT-STM/STS) investigations were conducted at the Laboratorio de Microscopías Avanzadas of the Universidad de Zaragoza, with a Scienta Omicron q-plus microscope, cooled down to 4.8 K, using a W tip, which for some measurements was functionalized with CO. LT-STS spectra grids and differential conductivity maps at constant height were obtained through an internal lock-in amplifier with an oscillation frequency of f$_{osc}$ = 817.3 Hz and a modulation amplitude of V$_{RMS}$ = 20 mV. %RT-STM measurements were performed at NanoLab of Politecnico di Milano, using home-made electrochemically etched tungsten tips, and at the Surface Supramolecular Chemistry Lab at the Università degli Studi di Padova, using electrochemically etched Pt-Ir tips.
XPS measurements were performed in situ at room temperature (RT) in UHV conditions, using a VG Scienta XM 650 X-ray source. The emitted radiation was monochromatized with a VG Scienta XM 780 monochromator optimized for Al K$\alpha$ radiation (1486.7 eV). Photoelectrons were collected in normal emission geometry and analyzed using a Scienta SES 100 electron analyzer integrated into the STM preparation chamber. 

STM data were analyzed using the image processing Gwyddion 2.63 software \cite{nevcas2012gwyddion}.
STS data were analyzed using Igor Pro and SpectraFox.
XPS spectra were fitted using IGOR Pro 6.37. Prior to fitting, the background contribution of the Br 3d region including the Au 5p$_{1/2}$ shoulder and Fe 2p$_{3/2}$  was subtracted. The binding energy scale was calibrated by aligning the position of the Au 4f peaks. The metallic Fe 2p core-level spectra were fitted using a Doniach-\v{S}unji\'{c} pair function, which accurately describes the asymmetric line shape of metallic photoemission peaks.

\subsection*{Theory}

Theoretical calculations are based on Density Functional theory as implemented in the SIESTA code \cite{Sole02}, i.e. by exploiting a pseudopotential description of electron-ion interaction and atomic orbitals basis set. A mesh cutoff of 400 Ry and a GGA-PBE \cite{PBE} exchange and correlation functional have been adopted.  The van der Waals correction to simulate the interaction between the covalent network and the Au substrate was included through Grimme dispersion forces \cite{Grimme}.
The supercell was built according to the experimental observation, i.e. with the covalent network forming an angle of 9$^\circ$ with respect to the [11$\overline{2}$] direction. The covalent network has been adapted to the Au(111) supercell built with the theoretical equilibrium lattice constant of 0.297 nm, resulting in a negligible strain of 0.7\% of the covalent system with respect to the freestanding case \cite{rabia2020structural}. The covalent network and the first layer of the surface have been relaxed until forces reached the convergence criteria of 0.04 eV/\AA.   
A mesh of $4\times4\times1$ k-points has been used for self consistency and a five times larger mesh for the PDOS. 
Calculations with HSE06 hybrid functional have also been performed using the relaxed structure obtained with GGA-PBE functional.
LDOS maps were obtained through calculations that exploit plane wave basis set, as implemented in Quantum Espresso. \cite{QE}
STM images are generated in the Tersoff-Hamann \cite{Tersoff} approach at constant distance or constant current, by integrating states in an interval [-0.5,E$_F$]. The STS maps were obtained by integrating the LDOS in a small interval of 0.1 eV at the fixed bias. To reproduce the effect of the experimental feedback, the maps are taken at a fixed distance z(x,y), where the lateral dependence of the tip height is set by realizing a constant current topography (current 1$\times 10^{-6}$), performing an energy integration of the electronic density, from the peak position to the Fermi level. 
%Simulated dI/dV maps have been obtained from a setpoint constructed by integrating the LDOS from the energy of the corresponding peak up to the Fermi level and selecting a constant isovalue of the integrated LDOS.

\section{Data availability}
The data supporting the findings of this study are available in the Article and its Supplementary Information. Additional data are available from the corresponding authors on request.

\bibliography{sn-bibliography}% common bib file

\section{Acknowledgments}
A.C., A.L.B. and C.S.C acknowledge funding by: Funder: project funded under the National Recovery and Resilience Plan (NRRP), Mission 4 Component 2 Investment 1.3 - Call for tender No. 341 of 15.03.2022 of Ministero dell'Universit\'a e della Ricerca (MUR); funded by the European Union NextGenerationEU - Award Number: project code PE0000021, Concession Decree No. 1561 of 11.10.2022 adopted by Ministero dell’Universit\'a e della Ricerca (MUR), CUP D43C22003090001, Project title ``Network 4 Energy Sustainable Transition NEST' '. F.S. acknowledges the financial support from the University of Padova through the grant  P-DiSC\#02BIRD2024-UNIPD  VS-PpyTM. J.L.C. acknowledges financial support from the Spanish Ministry of Science and Innovation (Grant PID2022-138750NB-C21 and ``Severo Ochoa' ' Programme for Centres of Excellence in R\&D CEX2023-001286-S funded by MCIN/AEI/ 10.13039/501100011033 and by ERDF ``A way of making Europe' ' ), from the regional Governments of Aragon (E12\_23R) and from the European Commission (project ULTIMATE-I, grant ID 101007825). J.L.C. further acknowledge the use of Servicio General de Apoyo a la Investigaci\'on-SAI and the Laboratorio de Microscopías Avanzadas of the Universidad de Zaragoza. S.A. acknowledges the CINECA award under the ISCRA initiative, for the availability of high-performance computing resources and support, and the financial support from INFN (Progetto iniziativa specifica ``TIME2QUEST''). A.O.B. acknowledges support from the Ministero dell’Universit\'a e della Ricerca (MUR) and the University of Pavia through the program ``Dipartimenti di Eccellenza'' 2023--2027.

\section{Author contributions}
A.C. conceived and conducted the STM experiments, analyzed the corresponding data, and wrote the initial draft. S.A. conceived the theoretical modelling, performed the DFT calculations, analyzed the corresponding data and wrote the initial draft. A.C. and G.C. performed preliminary RT-STM measurements at University of Padova.
A.C., E.P.S. and J.L.C. conducted LT-STM/STS experiments. J.L.C. contributed to the LT-STS data analysis and interpretation. A.O.B. synthesised the tBEB molecular precursor. J.L.C., F.S., F.T., A.L.B. contributed to the supervision and discussion of results. C.S.C. conceptualized, coordinated and supervised the work.  A.C. and S.A. contributed equally to this work. All authors contributed to the revision and final discussion of the manuscript.

\section{Competing interests}
The authors declare no competing interests.



\section*{Supplementary material}

## Br overlayer and h-GDY orientation on Au(111)

On Au(111), chemisorbed Br atoms deriving from the dehalogenation of tBEB molecules arrange according to $\sqrt{3}\times\sqrt{3}$ reconstruction, rotated by 30° with respect to the surface [8, 10]. In Figure S1a the porous network formed by Br atoms for a high-coverage tBEB sample is reported, displaying locally agglomerates with the $\sqrt{3}\times\sqrt{3}$ arrangement (close-up in Figure S1b)

Figure S1d shows the 2D FFT of the system reported in Figure S1c, from which the periodicity of the metalated system can be extracted, perfectly matching the one extracted from line profile measurements over atomically-resolved LT-STM images of the metalated system, i.e. 2 nm. LEED measurements allow to determine the 0° rotation angle of the metalated h-GDY with respect to Au(111) high symmetry directions: indeed, from the LEED pattern shown in Figure S1f, a (7×7)R0° supercell for the metalated h-GDY over the Au(111) lattice (of constant $a_{Au} = 0.288$ nm) can be estimated, from which the OMF periodicity is calculated at 2.016 nm. In particular, the yellow dashed hexagon corresponds to the first-order spots of the Au(111) lattice, and the red one corresponds to the metalated h-GDY. The LEED pattern reported in Figure S1g is simulated accordingly, through the commensurate superlattice matrix $\begin{pmatrix} 7.00 & 0.00 \\ 0.00 & 7.00 \end{pmatrix}$, and closely reproduces the experimental result.

2D FFT images were also employed to extract the periodicity of the extended $FeBr_2$ regions, i.e., as exemplified in Figure S1h-i. Finally, the 2D FFT was employed to determine the periodicity of the covalent h-GDY (Figure S1l-m), resulting 1.57±0.02 nm, in agreement with the one extracted from line profile measurements over atomically-resolved LT-STM images.

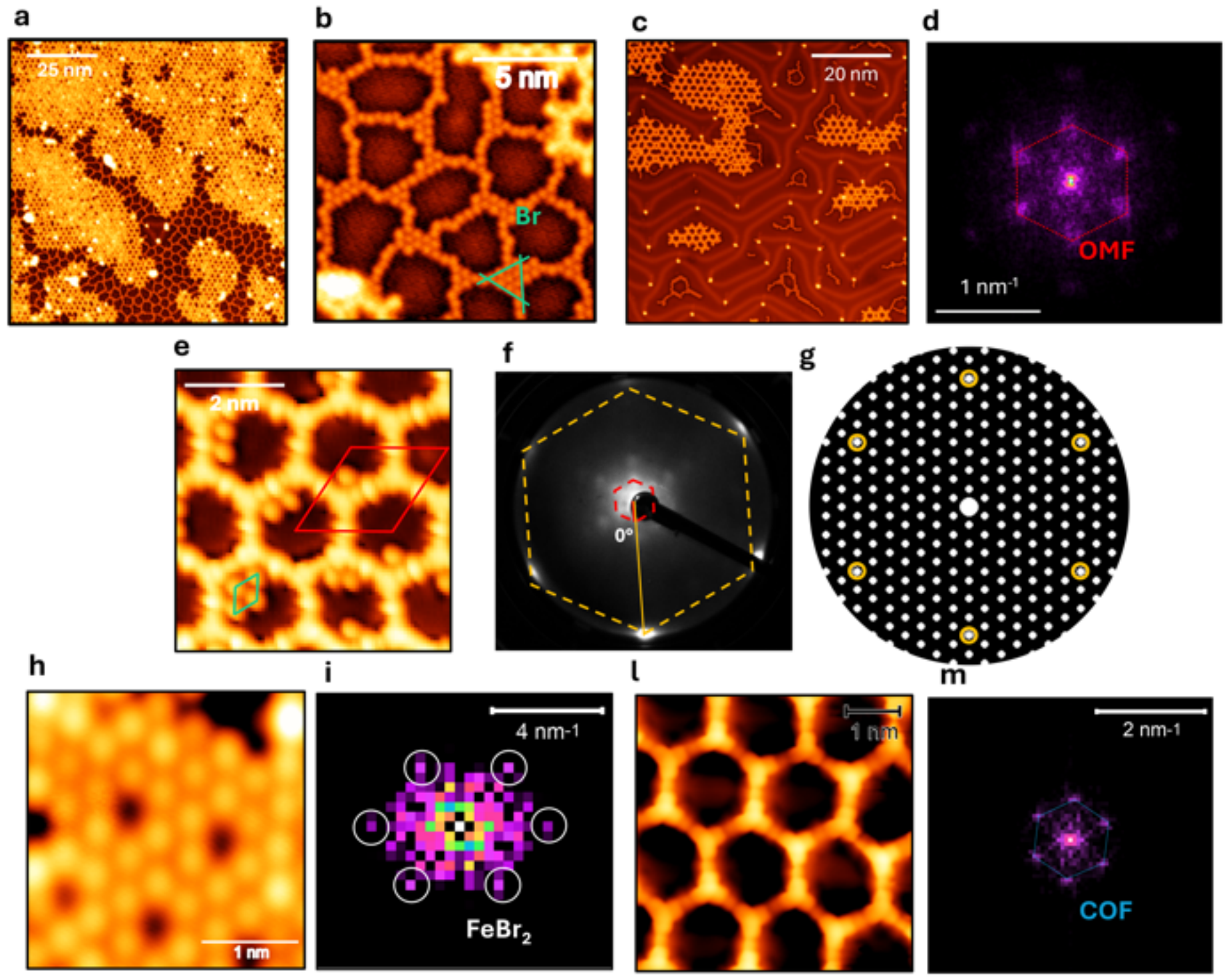


**Figure S1**: a) Large-scale LT-STM image of a high-coverage tBEB/Au(111) sample. The brighter network represents the as-deposited metalated h-GDY, while the darker network represents the $(\sqrt{3}\times\sqrt{3})$R30° Br reconstruction. b) Atomically resolved close-up of the $(\sqrt{3}\times\sqrt{3})$R30° Br reconstruction outside the metalated h-GDY on Au(111). c) Large-scale LT-STM image of a low-coverage tBEB/Au(111) sample, with isolated portions of the $(\sqrt{3}\times\sqrt{3})$R30° Br reconstruction visible on Au(111). d) 2D FFT of (c). e) Atomically resolved close-up of the metalated h-GDY, with a pore containing 4 Br atoms arranged according to the $(\sqrt{3}\times\sqrt{3})$ reconstruction. The red rhombus indicates the unit cell of the (7×7)R0° superlattice of the metalated h-GDY. The turquoise rhombus indicates the unit cell of the $(\sqrt{3}\times\sqrt{3})$R30° overlayer of Br atoms. f) LEED diffraction pattern of the metalated h-GDY on Au(111) (setpoint: 42 eV) for a high-coverage sample (∼0.85 ML). The yellow hexagon corresponds to Au(111) first-order LEED spots, while the red hexagon to the metalated h-GDY. g) Simulation of the metalated h-GDY commensurate LEED pattern through the superlattice matrix $\begin{pmatrix}7.00 & 0.00\\ 0.00 & 7.00\end{pmatrix}$; the first-order spots of Au(111) are marked in yellow. h) Atomically-resolved hexagonal structure of $FeBr_2$ on Au(111). i) 2D FFT of (h). White circles highlight the spots used to extract the $FeBr_2$ periodicity. l) Atomically resolved LT-STM image of the covalent h-GDY. m) 2D FFT of (l). The spots corresponding to the covalent h-GDY are highlighted by the blue dashed hexagon. STM set-point: a) -1V/5pA, b) -1V/5pA, c) -500mV/20pA, e) -50mV/10pA, (h)-3mV/30pA, (l) -4mV/100pA.

## XPS - C1s, Br3d and Fe2p

XPS analyses were performed following Fe deposition on tBEB/Au(111) samples to investigate the catalytic role of iron in the Ullmann-like coupling reaction.

The Fe $2p_{3/2}$ spectrum after annealing at 385 K, at which the transition to the covalent phase is already ongoing, shows a shift toward higher binding energy compared to metallic Fe (centered at 707 eV [1, 5]), with a main peak at approximately 709.5 eV and a satellite feature around 715 eV. The fitting was performed following the literature for submonolayer $FeBr_2$ on Au(111) [4], where the main peak at 709.5 eV is assigned to $Fe^{2+}$ species, while the satellite at 715 eV is associated with final-state effects related to the multiplet structure of the 2p core level in transition metals. In the Br 3d region, the Br $3d_{5/2}$ component, initially centered at 68 eV for bromine adsorbed on Au [3, 2], shifts to 69 eV upon Fe deposition, consistent with Br bound in Fe-Br coordination environments. This shift is in line with literature reports for $FeBr_2$ on Au(111) [4]. Temperature-dependent XPS measurements reveal that, upon annealing at 570 K, the Br 3d signal is strongly suppressed and nearly disappears. At the same time, the Fe 2p signal shows a marked reduction of the $FeBr_2$-associated component, indicating that the Fe–Br interaction is no longer stable at elevated temperature. The simultaneous disappearance of bromine and the modification of the iron spectral features suggest that, at about 570 K, all bromine desorbs from the surface, while iron is no longer bound in a stable $FeBr_2$ phase. A fraction of Fe may remain on the surface as metallic adatoms or residual $FeBr_2$, while another fraction may diffuse into subsurface sites of the Au(111) substrate, as reported in the literature [6].

The evolution of the C 1s spectra as a function of temperature reveals significant changes both in the binding energy position and in the overall peak shape (Figure S3), indicating a progressive modification of the chemical environment of carbon atoms during the reaction. After deposition of iron and annealing at 385 K, the C 1s spectrum is characterized by a multicomponent structure, including a distinct low-binding-energy (BE) contribution at 283.48 eV (highlighted in green in Figure S3), assigned to the $C_4$ carbon atoms, involved in metalated C–Au bonds. This assignment is consistent with previous reports [11, 9, 2], where such low-BE features are associated with carbon atoms coordinated to Au adatoms on the substrate. Upon annealing at 420 K, a clear spectral evolution is observed. The low-BE component attributed to C–Au bonding disappears, leading to a shift of the overall C 1s envelope toward higher binding energies. This shift reflects the loss of gold–carbon interaction and the transition to a purely covalent carbon network.

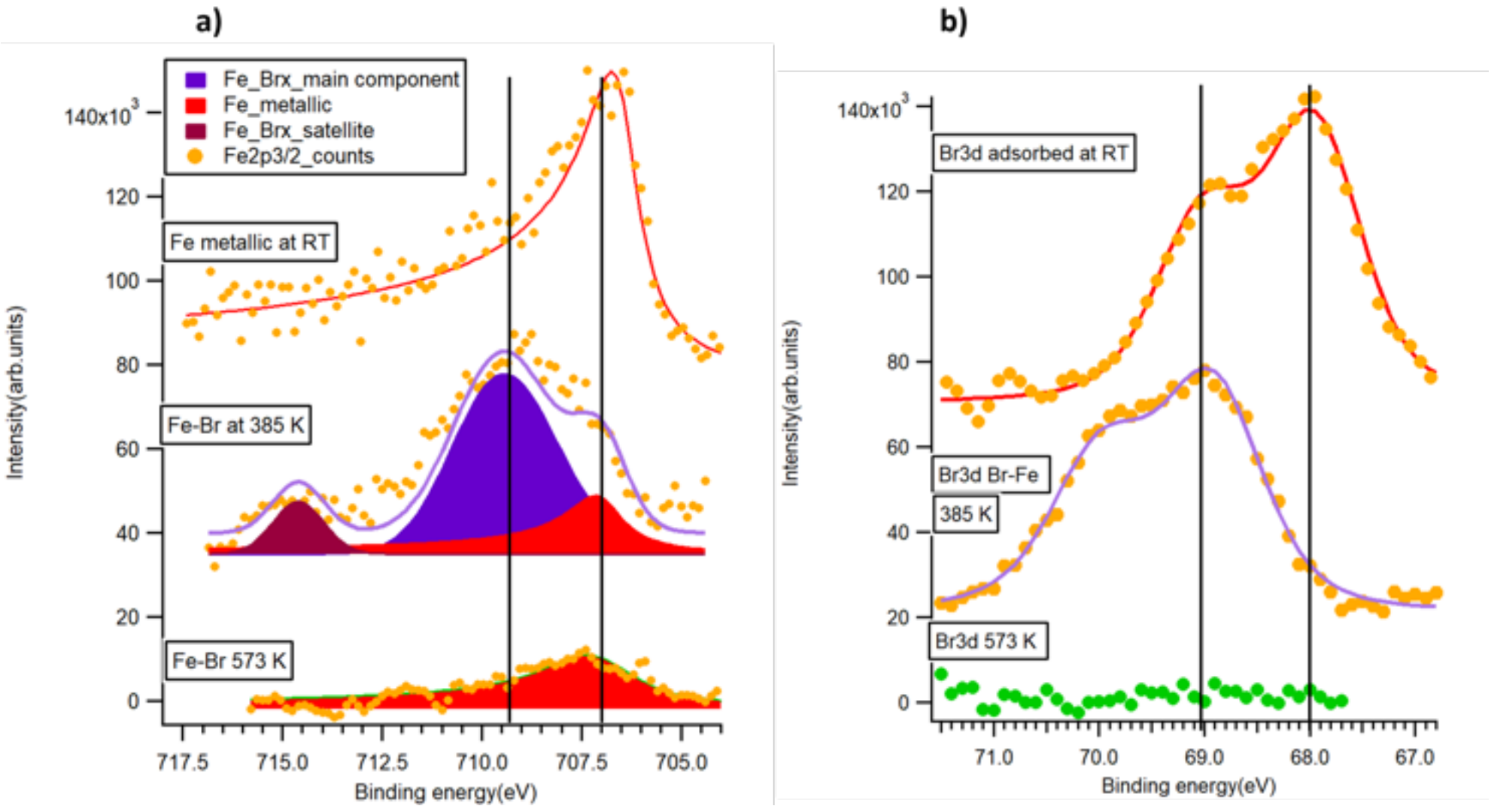


**Figure S2**: Temperature-dependent XPS spectra of: (a) Fe $2p_{3/2}$. The spectra are deconvoluted into components corresponding to different chemical environments of Fe. The red component is assigned to Fe species adsorbed on the Au(111) surface. Upon annealing at 385 K, it can be observed also the presence of purple components, attributed to Fe and Br atoms bonded in the $FeBr_2$ structure. (b) Br 3d core levels. The red line corresponds to bromine adsorbed on the Au surface at RT, the purple line to bromine bonded to iron at 385 K, and the green line to negligible residual bromine traces on the surface at 573 K. Dots correspond to the experimental data, while the solid lines to the fitted components.

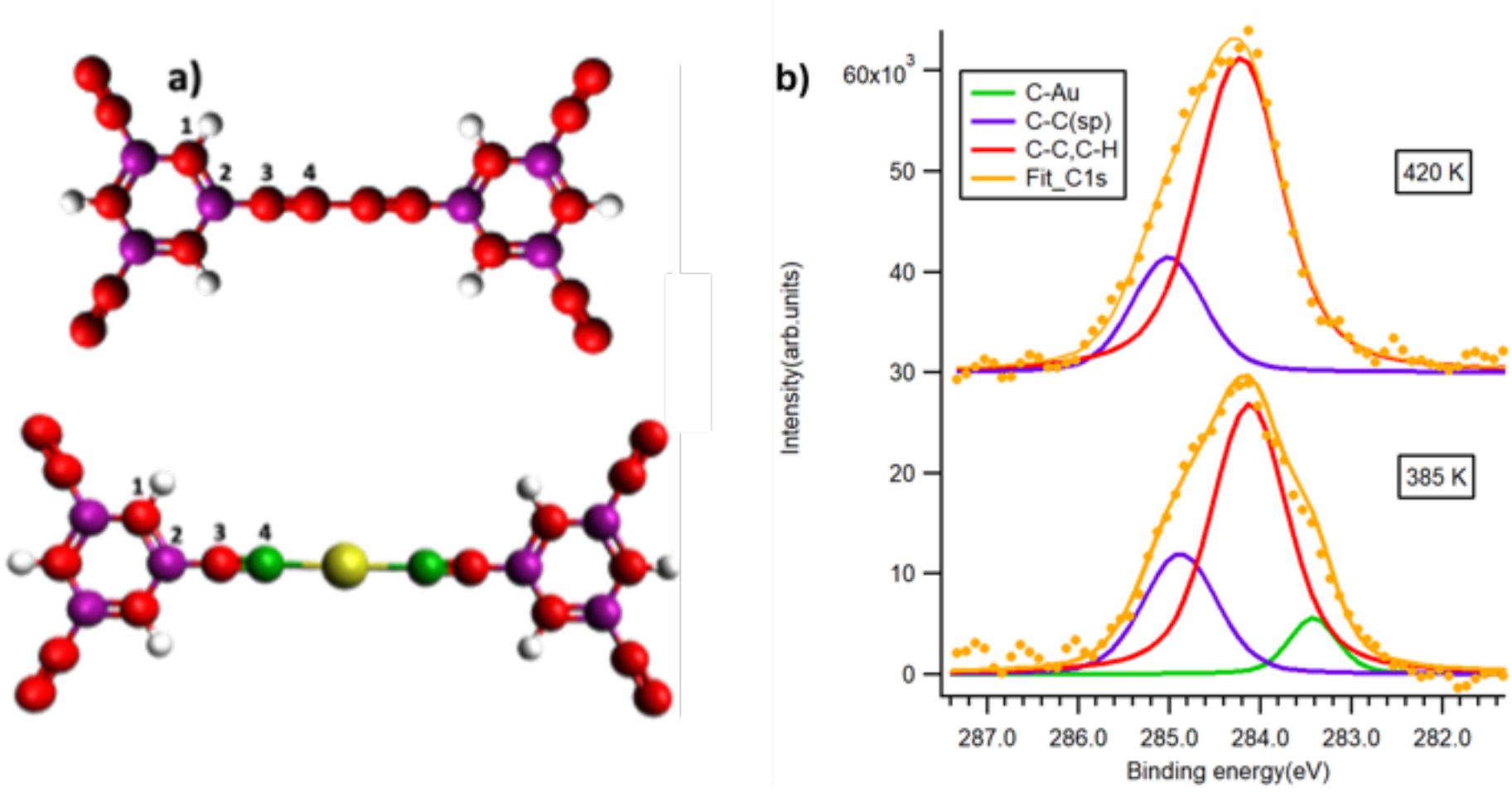


**Figure S3**: (a) Ball-and-stick atomic models of the metalated (bottom) and covalent (top) h-GDY dimer. The different colors identify distinct carbon chemical environments: red corresponds to C–C/C–H carbon atoms, purple to C–C carbon atoms bonded to the acetylenic group, and green to C atoms bonded to Au. Au adatom is highlighted in yellow. (b) Fitted C1s XPS spectra after Fe deposition and annealing at 385 K (bottom) and 420 K (top).

## Fe-mediated OSS: 420 K annealing

Figure S4 shows LT-STM images of the system after annealing at 420 K.

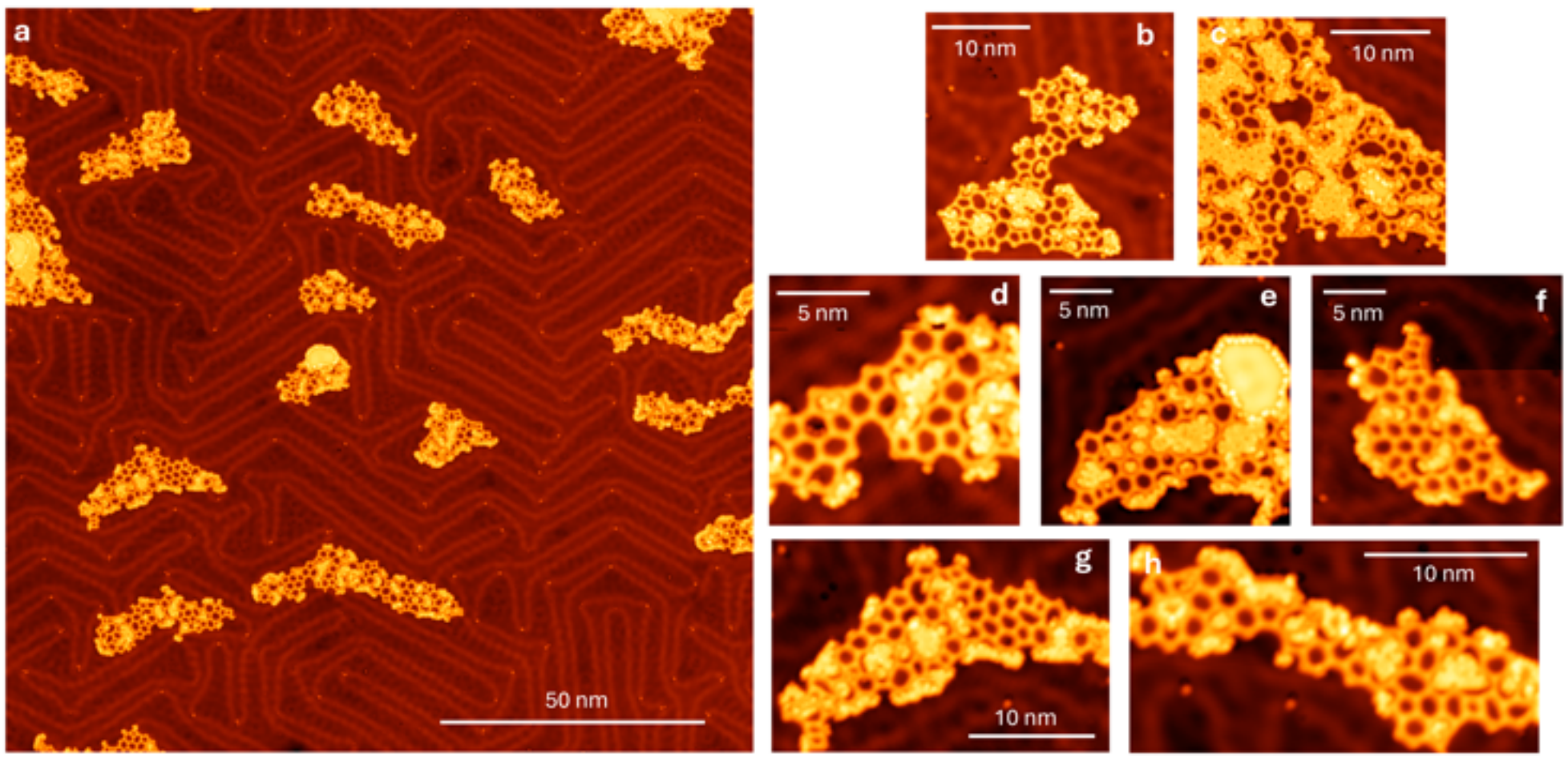

**Figure S4**: Large-scale and close-up LT-STM images of covalent h-GDY islands after the annealing at 420 K. Bright areas correspond to $FeBr_2$. STM set-point: (a) -2mV/2pA, (b-c) -2mV/10pA, (d)-2mV/2pA, (e)-2mV/140pA, (f-h) -2mV/2pA.

## Fe-mediated OSS: 610 K annealing

Figure S5 shows LT-STM images of the system after annealing at 610 K.

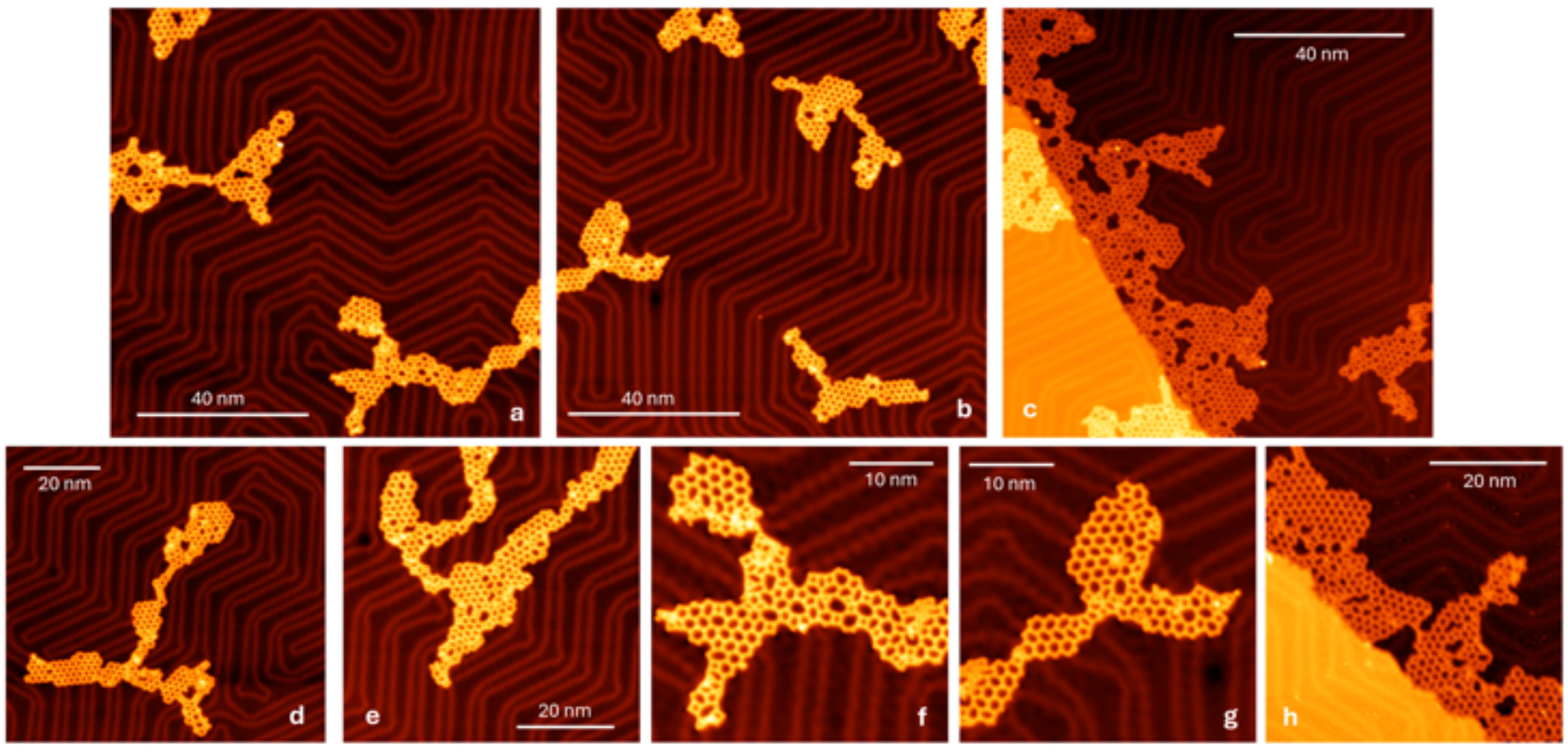


**Figure S5**: Large-scale LT-STM images of covalent h-GDY islands after the annealing at 610 K. STM set-point: (a-e) -1V/5pA, (f-g) -5mV/100pA, (h) 100mV/50pA.

## Br adatoms stability

We performed DFT calculations for the metalated h-GDY with one adsorbed Br atom in different positions with respect to the underlying substrate, i.e. TOP, FCC, HCP. The FCC adsorption site is the most stable with a total energy gain of 63 meV and 68 meV with respect to HCP and TOP adsorption, respectively. We also estimated the binding energy of Br adatoms, as:

$$E_{ads} = E_{Br/OMF} - E_{OMF} - nE_{Br},$$

where $E_{OMF}$ is the energy of metalated h-GDY without adatoms, $n$ is the number of Br atoms adsorbed and $E_{Br}$ is the reference energy of a Br atom in the gas phase. The adsorption energy of a single Br atom is $-3.99$ eV, while the subsequent adsorption of a second and third adatom costs 19 meV and 77 meV, respectively. For a number

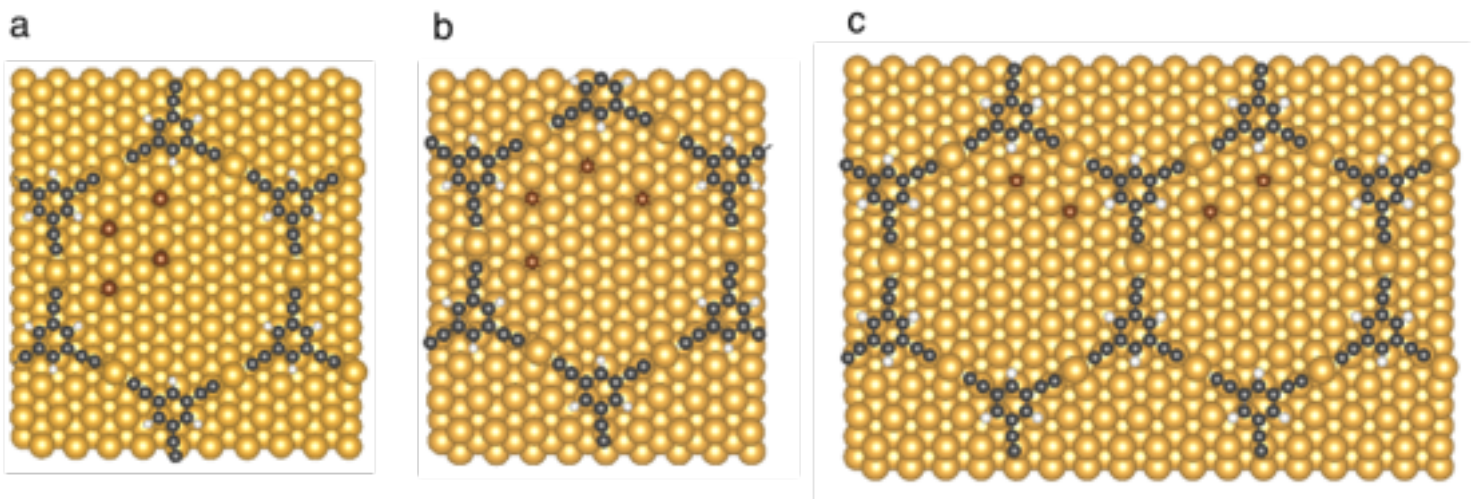


**Figure S6**: Structural model of a Br cluster in the $\sqrt{3} \times \sqrt{3}$ arrangement (a), linear (b) and "2+" distributed (c) configurations of four Br atoms in a pore of the metalated h-GDY structure.

of Br atoms in the pore larger than three, two possible arrangements were considered: the $\sqrt{3} \times \sqrt{3}$ like arrangement (see Figure S5a) and a "linear" configuration with all the Br atoms along the edge of the pore (Figure S6b). The calculation for 4 Br adatoms predicts a larger stability for the former, with an energy gain of 54 meV. It is worth noting that, anyway, the alternative configuration with four Br atoms distributed in adjacent pores, i.e. "2+2" in Figure S6c, is 170 meV more stable than the $\sqrt{3} \times \sqrt{3}$, demonstrating the preference of Br to distribute in different pores. This energy difference is calculated as:

$$E_{4Br} = E_{4\ linear} - E_{2+2} - E_{OMF}$$

This tendency is confirmed when the number of Br is even larger: for six Br adatoms, the energy gain of distributing in different pores (e.g "2+4" or "3+3") is in the order of 0.5 eV. We also verified that Au adatom stabilizes Br adatoms at  5 Å distance. In particular, two Br adatoms are stabilized by the proximity of an Au adatom when they form an equilateral triangle with all the adatoms in FCC positions (Figure S7c).

If the mid bondlength point is instead displaced near the adatom or shifted laterally, as in Figure S7a,b, the total energy increases by 32 meV and 38 meV, respectively. Positioning the Au adatom far apart increases the energy even more: in the configuration shown in Figure S7 c, the system is 96 meV less stable.

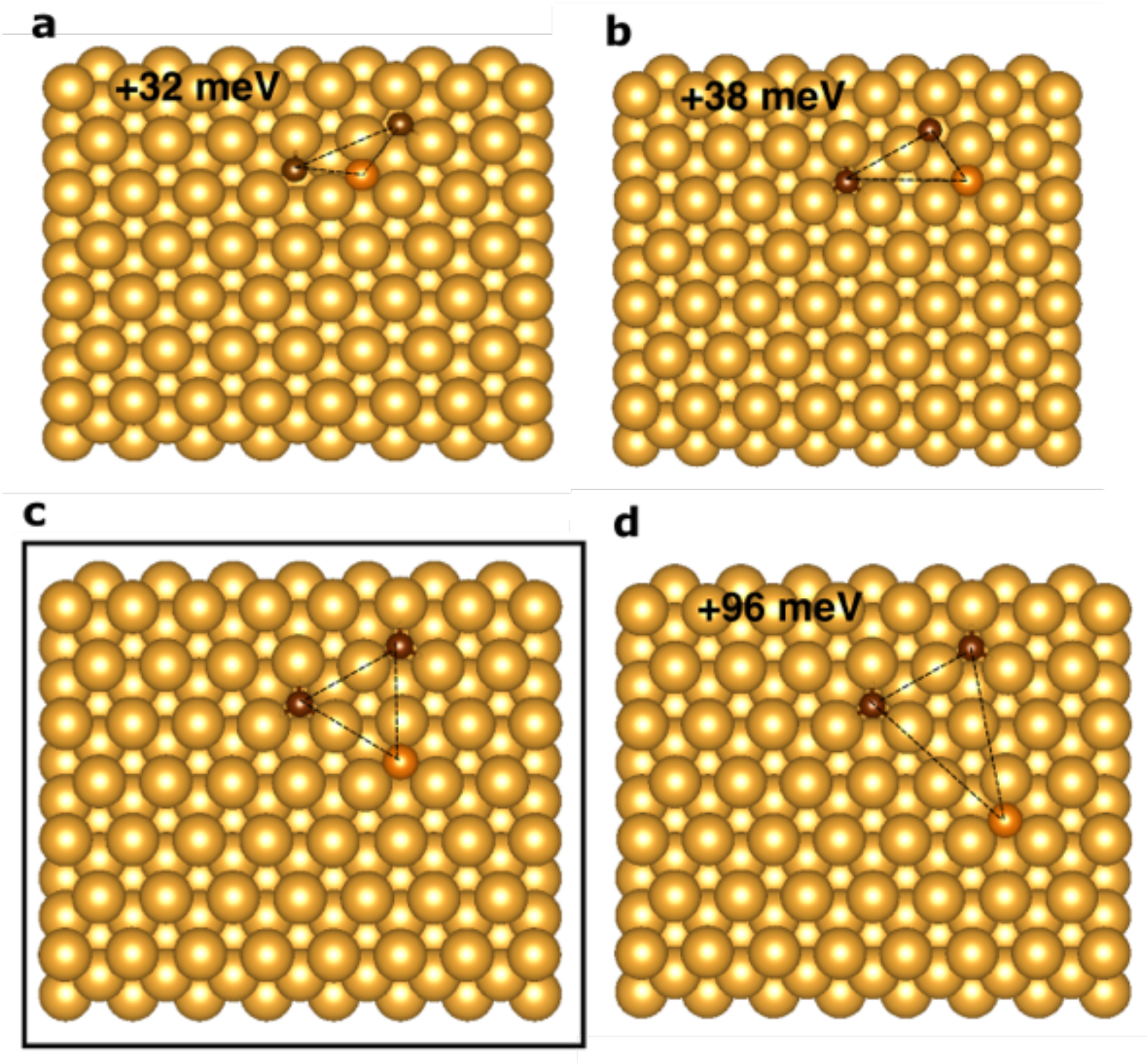


**Figure S7**: Different positioning of Au adatom (orange) with respect to two Br atoms (brown) and relative energies.

## Fe-Br interaction

We evaluate the strength of the Br-Fe interaction, mediated by gold, through different model calculations. We considered the case in which a Fe adatom is far apart from a Br adatom, i.e. in the center of the pore, while the Br atom is near the edge of the OMF (Figure S8a). We compare this situation with the case in which the Fe adatom stands near the Br (Figure S8b). We verified that the latter is 1.1 eV more stable than the former, leading to the formation of a Fe-Br dimer and confirming the preference of Br to bind to Fe with respect to substrate Au. Moreover, the FeBr dimer is stabilized by the proximity to a gold adatom, gaining 147 meV with respect to being located in the center of the pore, in the same adsorption configuration with respect to the substrate (Figure S8c). It is worth noting that when Br is bound to Fe, its distance from the substrate surface increases with respect to the single Br adatom, being farther than the Fe adatom in the dimer.

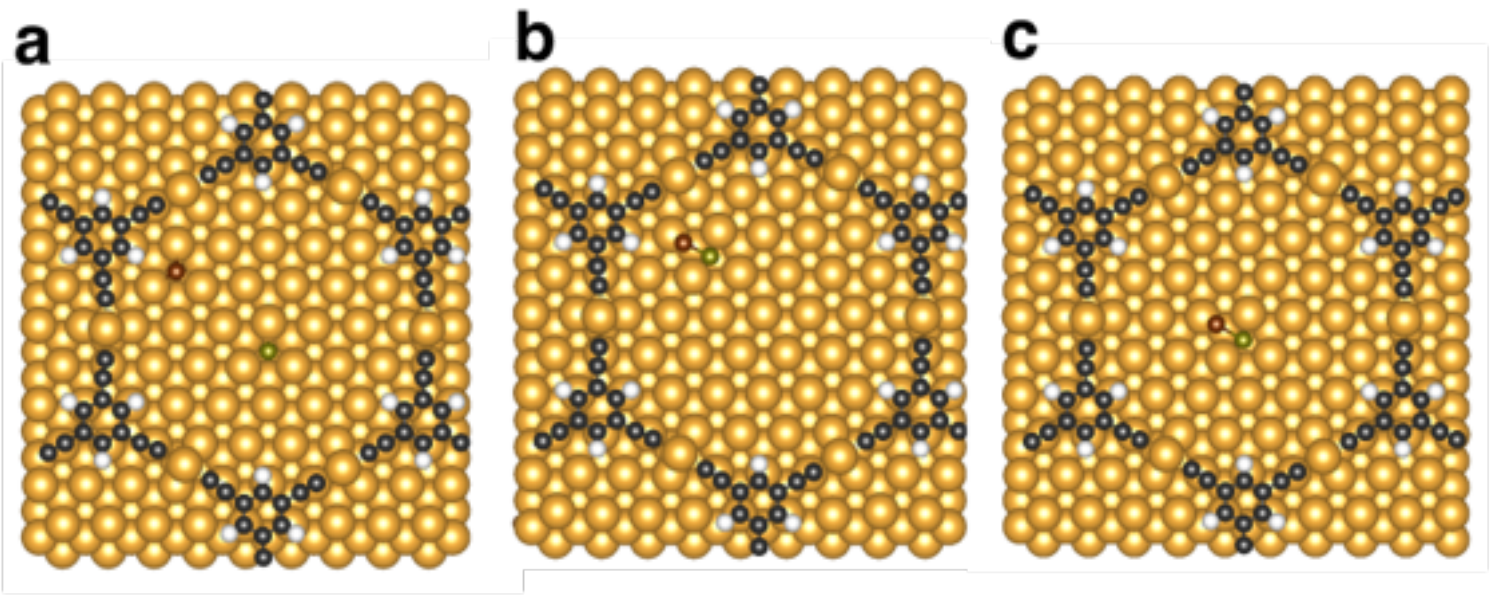


**Figure S8**: Different models for Br and Fe adatoms: not interacting (a), dimer near the h-GDY edge (b) and dimer far from the h-GDY edge (c).

We should note here that, apart from simple test purposes, the FeBr dimer would be quite unusual, with under-saturated Fe. For this reason, and in agreement with what was observed in the experiments, we repeated the calculations above considering 2 Br atoms and 1 Fe. First of all, we determined the preferential distance between the two species, as reported in Figure S9. Differently from Br-Au, Br and Fe tend to form a molecular complex with two Br atoms, with bond lengths of $\sim$ 2.2 Å. The binding energy of the molecule is $\sim$ 1.1 eV.

We also explore the energy landscape when the molecule is moved in different positions in the pore. Placing the molecule near the edge of the h-GDY implies an energy gain of 0.53 eV (Figure S10a,b). We also simulated a situation in which an Au adatom of one of the h-GDY edges is removed in the presence of a $FeBr_2$ molecule near the edge (Figure S10c) or far away (Figure S10d). We found that the process is 0.8 eV favoured in the former case.

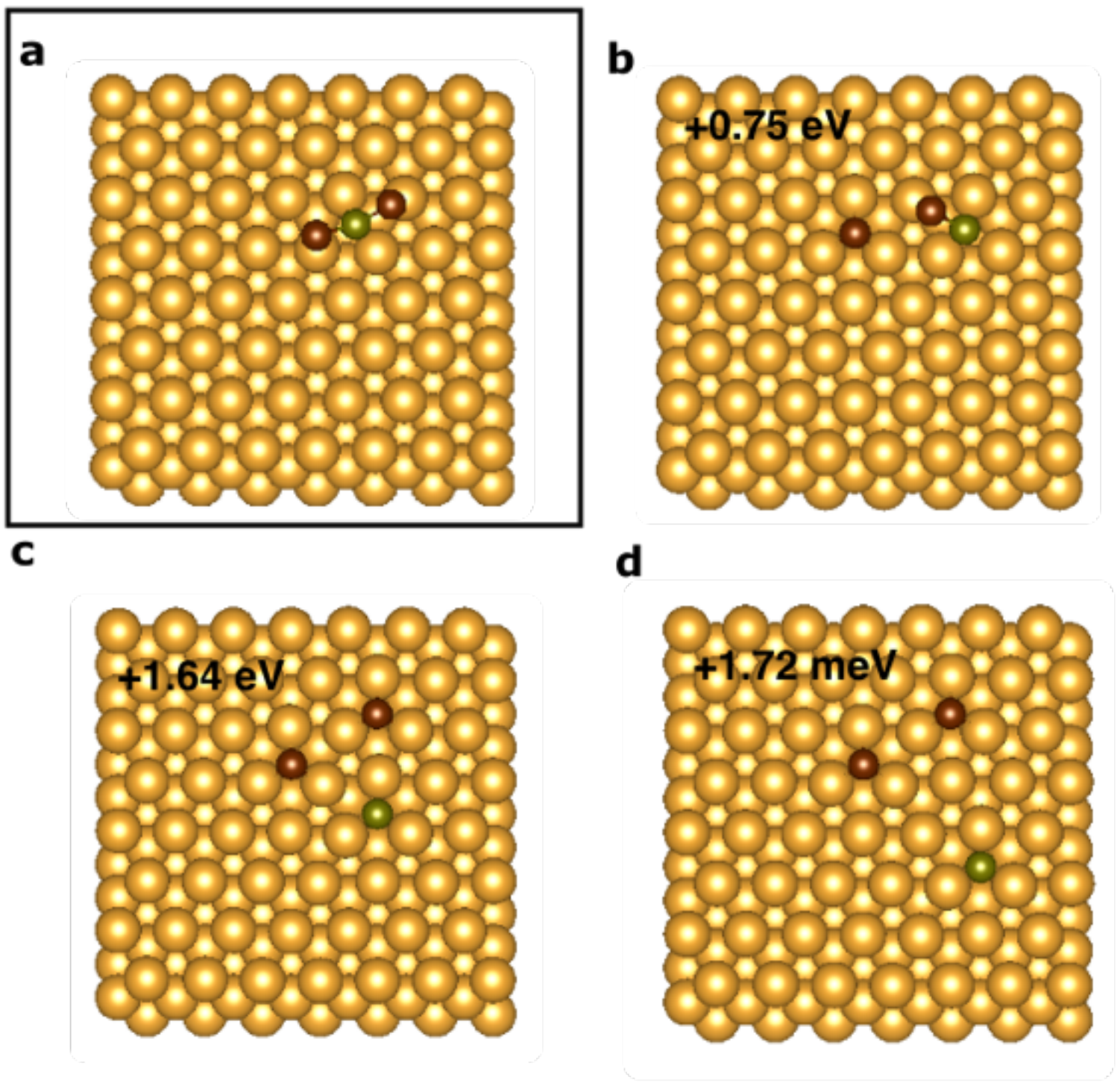


**Figure S9**: Different positioning of a Fe adatom (green) with respect to two Br atoms (brown) and relative energies (c).

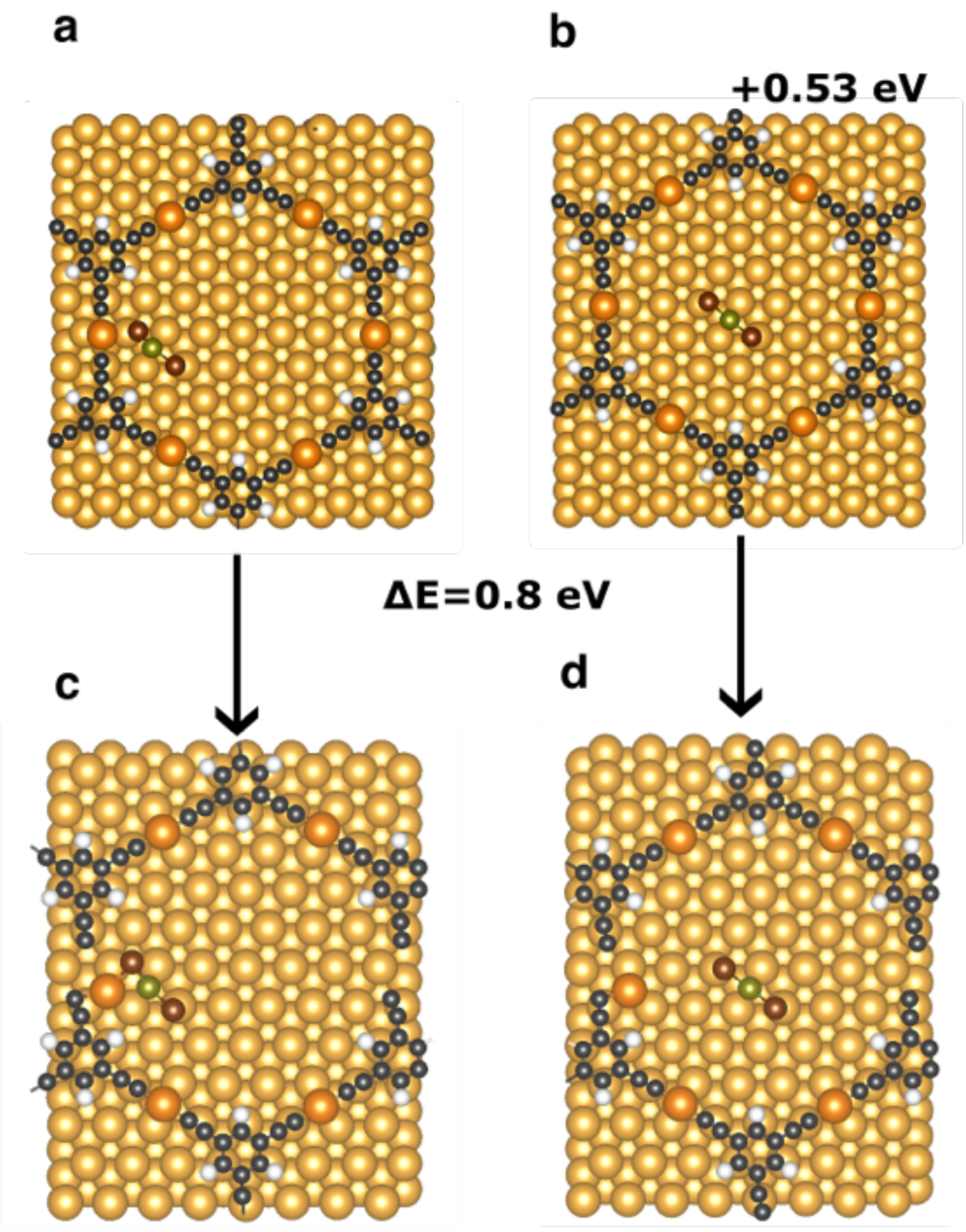


**Figure S10**: $FeBr_2$ molecule in different positions in the pore (a,b) and with the removal of an Au atom (c,d). Energy difference for the two locations of the molecule and for the two adatom removal processes is reported.

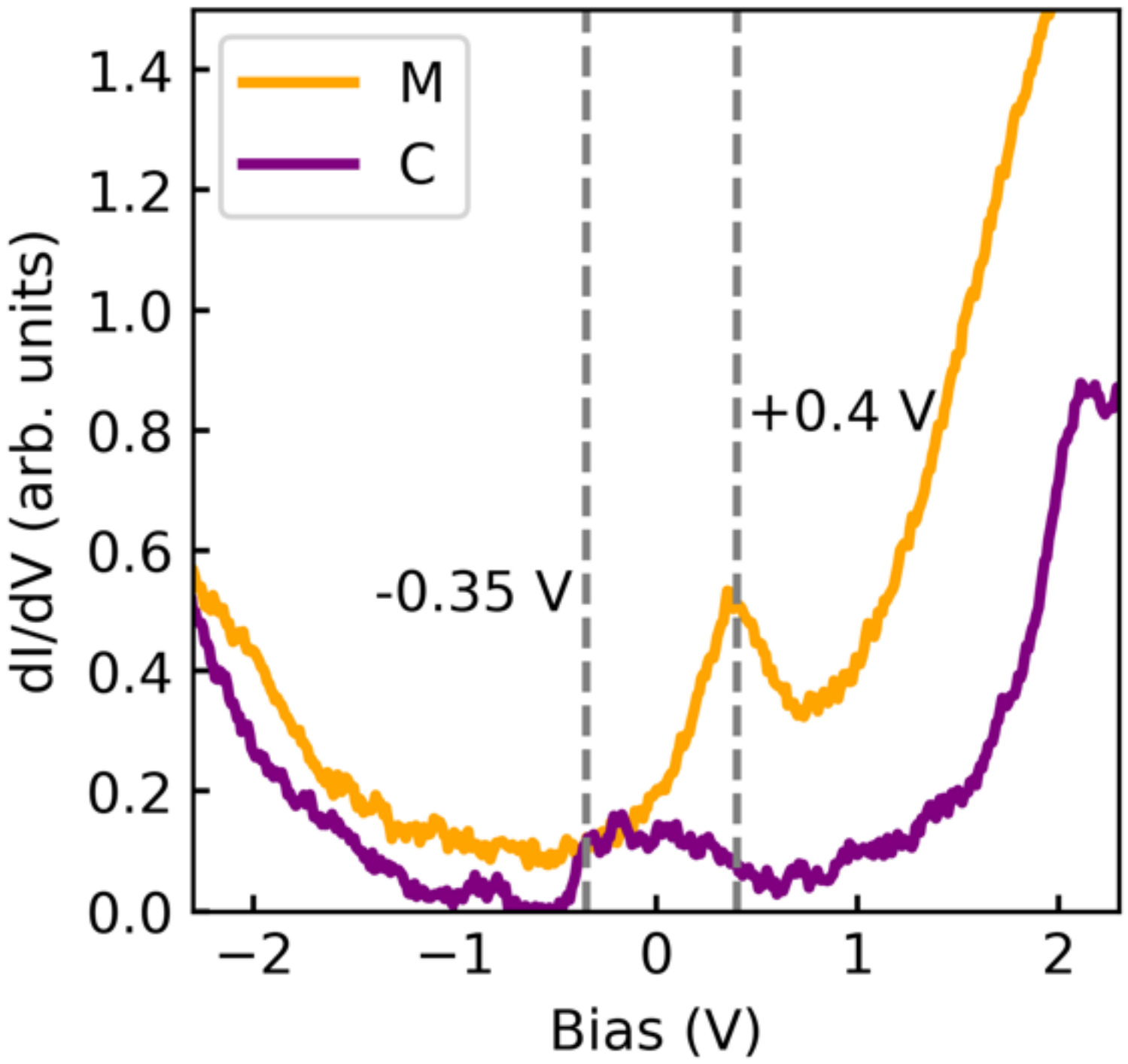


**Figure S11**: Conductance spectra of the Au(111) surface state in the middle of the pore of the metalated (M, orange) and covalent (C, purple) h-GDY network. Dashed vertical lines correspond to the energy position of the surface state, i.e. -0.35 V and +0.4 for the covalent and metalated system, respectively.

## STS and dI/dV maps of metalated h-GDY on Au(111)

The electronic properties of metalated and covalent h-GDY have been compared by means of STS and dI/dV measurements supported by the theory. The main features are:

- The shift of the spectral features toward higher energy in the metalated h-GDY compared to the covalent h-GDY. These features have been identified on the basis of the dI/dV maps.
- The reduction of the DOS intensity at the Fermi level passing from metalated to covalent h-GDY testifying a metal-to-semiconductor transition.
- The shift of the Au Shockley state, as discussed in the main text.

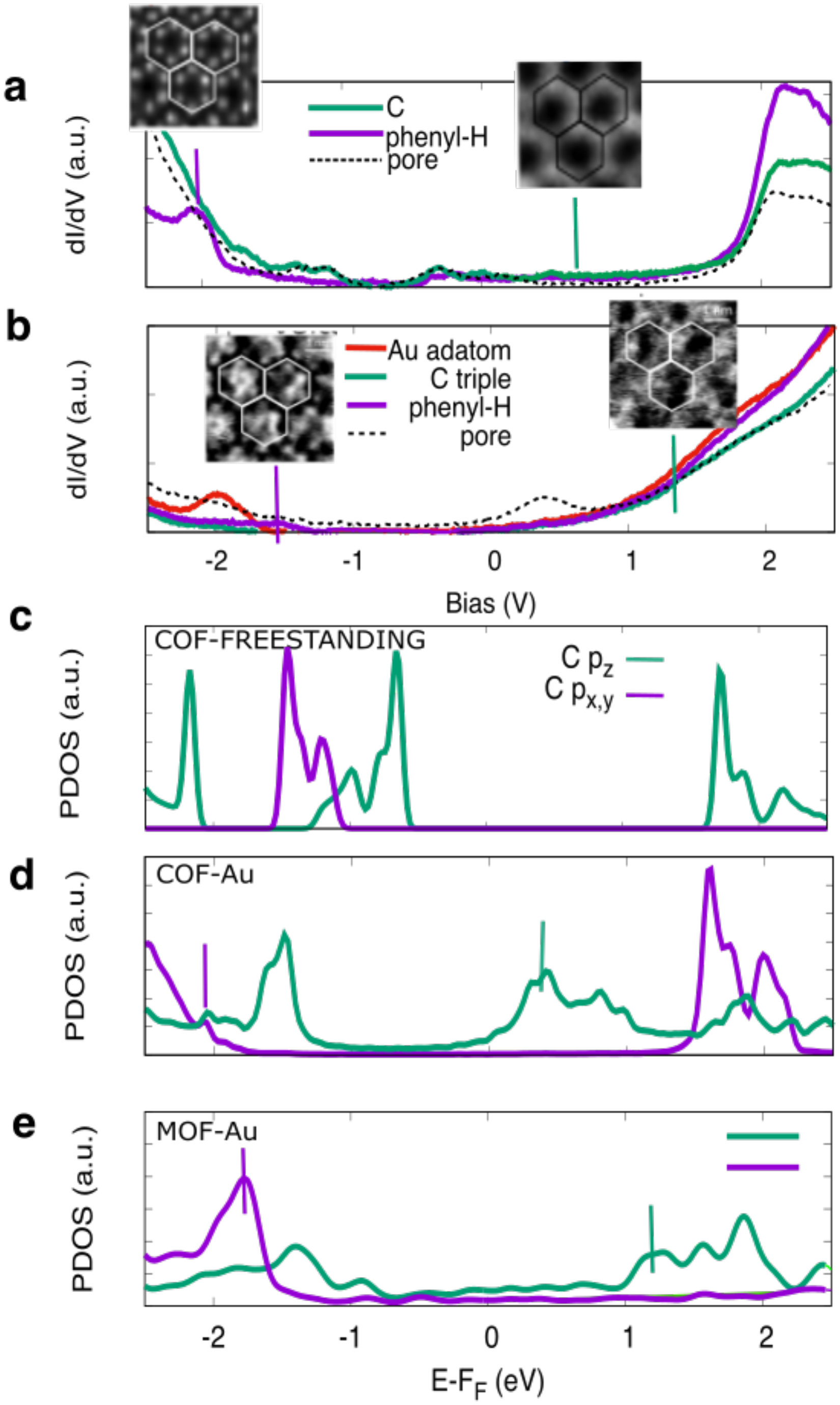


**Figure S12**: a) STS and dI/dV maps of covalent h-GDY on Au(111) in comparison with the STS of the metalated h-GDY (b). Purple vertical lines mark the bias at which dI/dV maps show the $p_x, p_y$ network. Green vertical lines mark the bias at wich dI/dV maps shows a $p_z$ network. c) orbital resolved PDOS of covalent freestanding h-GDY; d) and e) orbital-resolved PDOS of covalent and metalated h-GDY on Au(111), respectively. Vertical purple and green lines mark the position of $p_x, p_y$ and $p_z$ states that can be related to the ones reported in the experimental panels a) and b).

## Insight into LT-STS measurements

The p-type tip orbitals contributing to the tunneling current due to the presence of the CO-functional group allow for enhancing in-plane orbital features, while the mainly s-type orbital of the metallic STM tip enhances the visualization of the out-of-plane orbital features [7]. This explains why the dI/dV maps acquired with the CO-terminated tip allow for the better visualization of the orbitals at the H-terminated edges of the phenyl rings at -2.1 V and +2.2 V (predicted to be corresponding to in-plane carbon $p_{x,y}$ orbitals and H s orbitals, see the section “Electronic properties”), while the dI/dV maps acquired with the metallic tip allow for the better visualization of the orbitals along the covalent h-GDY network at -1.45 V and +0.5 V (predicted to be corresponding to out-of-plane carbon $p_z$ orbitals, see the section "Electronic properties”), as reported in Figure 3. Figure S13a-c shows that the dI/dV maps acquired with the CO-terminated tip at -1.45 V and +0.5 V allow as well for the visualization of C $p_z$ orbitals, even if with a lower resolution with respect to the metallic tip. In the same way, Figure S13d-f shows that the dI/dV maps acquired with the metallic tip at -2.1 V and +2.2 V allow as well for the visualization of C $p_{x,y}$ and H s orbitals, even if with a lower resolution with respect to the CO-terminated tip.

Interestingly, with both the CO-terminated and metallic STM tips, the acquired spectra show a much larger intensity at the energies corresponding to C $p_{x,y}$ and H s orbitals with respect to the energies corresponding to C $p_z$ orbitals. We relate this behaviour to the tip-sample distance during the spectral acquisition. Indeed, being a constant current mode measurement, the chosen setpoint that affects the tip-sample distance set as reference on the carbon network before starting the grid is the bottleneck for the type of orbital visualization during the whole grid. During the measurement, at each pixel of the scanned frame, the tip-sample distance is adjusted by the feedback loop to preserve the setpoint value of the tunnelling current. Once the tip-sample distance is fixed, the feedback loop is removed, and a bias sweep is performed, with the acquisition of the resulting tunnelling current as output. When the tip is scanned on top of the covalent network, i.e. on C sites (green dots in Figure 3), the contribution to the detected tunnelling current comes from p orbitals, and, in particular, by $p_z$ orbitals, which extend in the out-of-plane direction pointing towards the tip. When the tip is scanned inside the h-GDY pores (violet dots in Figure 3), i.e. on the H-terminated edges of the phenyl rings, the contribution to the tunnelling current comes from C $p_{x,y}$ orbitals and H s orbitals; thus, to reach the setpoint current, the tip must get closer. As a result, inside the h-GDY pores, in correspondence of the H-terminated edges of the phenyl rings, the spectra are acquired at a smaller sample-tip distance, resulting in highly intense features, i.e. the states found at -2.1 V and +2.2 V. Conversely, on top of the h-GDY network, the spectra are acquired at a larger sample-tip distance, resulting in weak features, i.e. the states found at -1.45 V and +0.5 V.

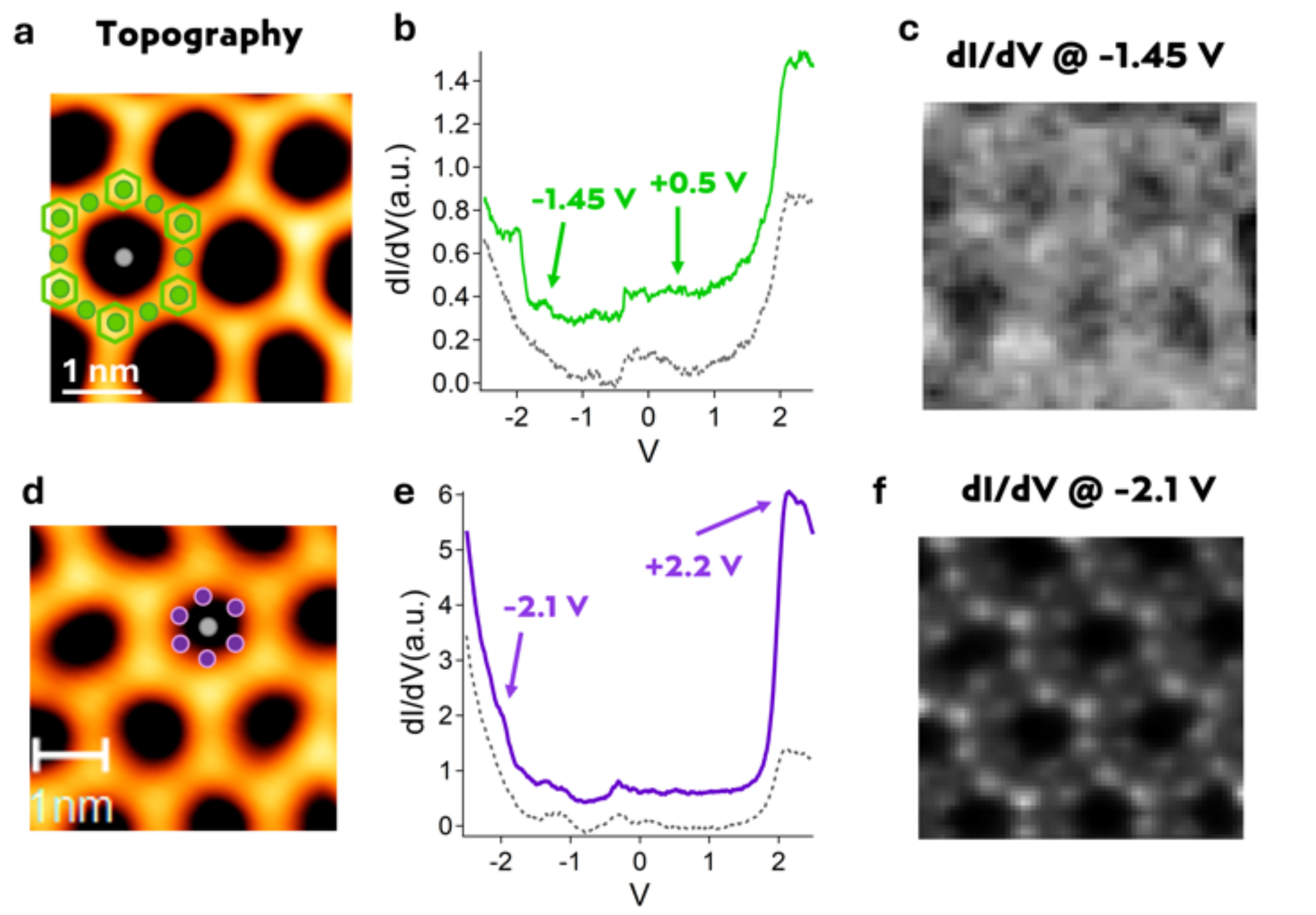


**Figure S13**: (a) Atomically-resolved LT-STM image of the STS grid area acquired with CO-functionalized tip. (b) LT-STS curves corresponding to the equivalently colored dots in (a). The grey dashed line corresponds to the spectrum acquired in the position of the grey dot in (a), i.e. the Au(111) surface state confined by the h-GDY pore. The green line corresponds to the average spectrum acquired in the position of the green dots in (a), i.e. the apexes and the arms of the hexagons forming the h-GDY, with the -1.45 V and +0.5 V spectral features highlighted by green arrows. (c) Constant-current dI/dV map acquired with CO-functionalized tip at -1.45 V. STS/STM set-point: -0.55 V/40 pA, with lock-in amplitude 20 mV and frequency 817 Hz. (d) Atomically-resolved LT-STM image of the STS grid area acquired with metallic tip. (e) LT-STS curves corresponding to the equivalently colored dots in (d). The grey dashed line corresponds to the spectrum acquired in the position of the grey dot in (d), i.e. the Au(111) surface state confined by the h-GDY pore. The violet line corresponds to the average spectrum acquired in the position of the violet dots in (d), i.e. the H-terminated edges of the phenyl rings of the h-GDY, with the -2.1 V and +2.2 V spectral features highlighted by violet arrows. (f) Constant-current dI/dV map acquired with metallic tip at -2.1 V. STS/STM set-point: -0.55 V/50 pA, with lock-in amplitude 20 mV and frequency 817 Hz.

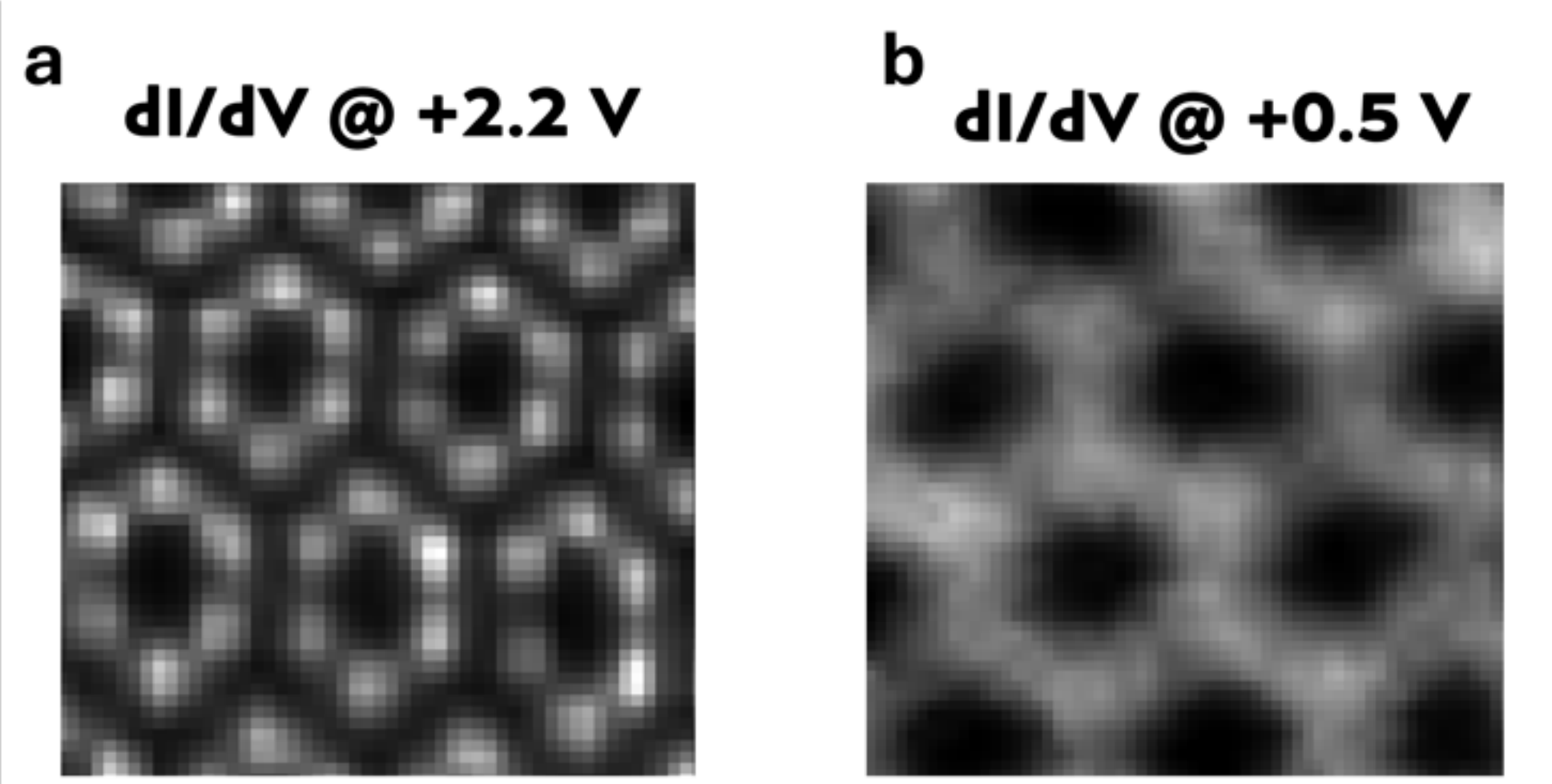


**Figure S14**: (a) Constant-current dI/dV map acquired with CO-functionalized tip at +2.2 V. STS set-point: -0.55 V/40 pA, with lock-in amplitude 20 mV and frequency 817 Hz. (b) Constant-current dI/dV map acquired with metallic tip at +0.5 V. STS set-point: -0.55 V/50 pA, with lock-in amplitude 20 mV and frequency 817 Hz.

## Density of States with hybrid functionals

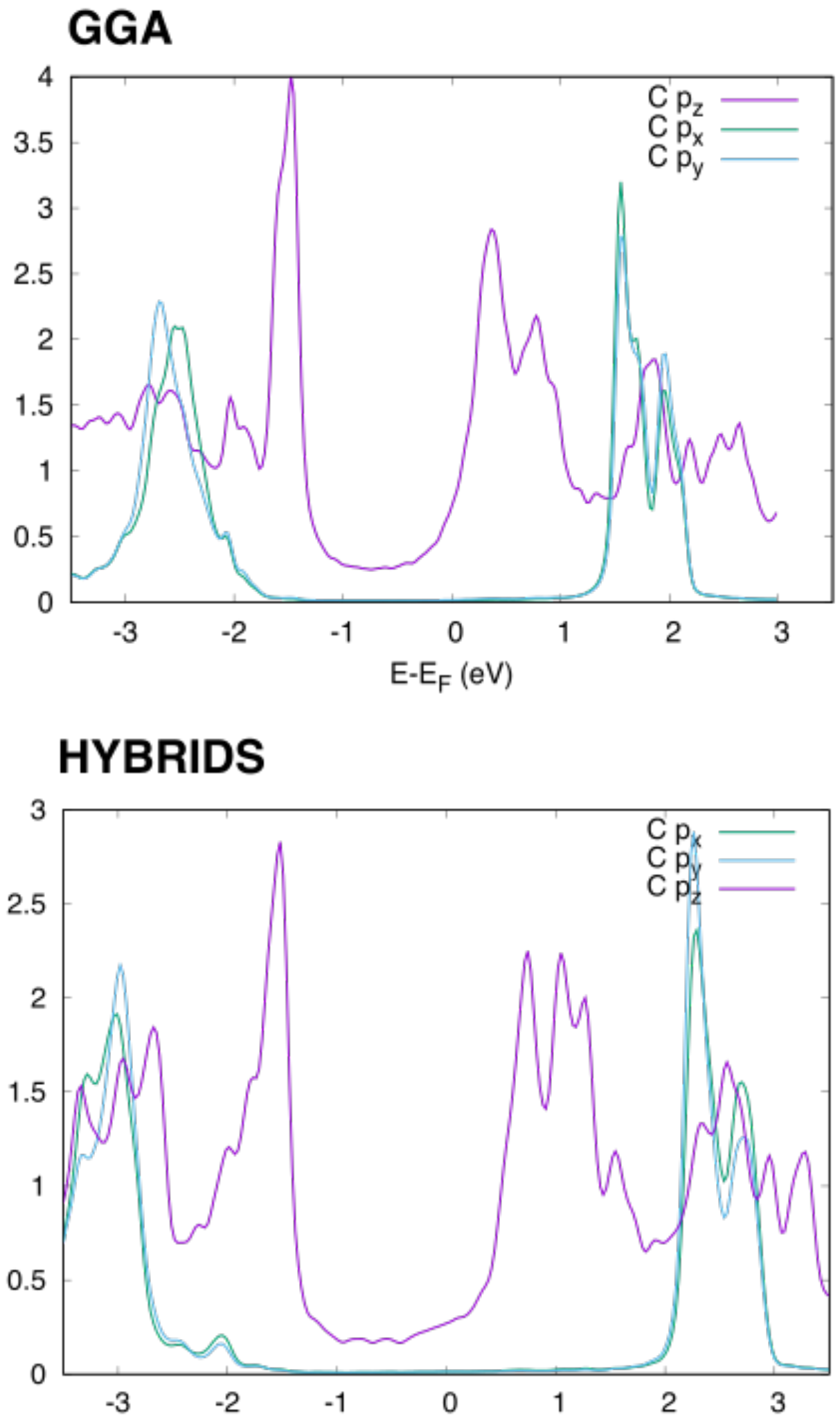


**Figure S15**: Comparison between the PDOS of carbon for covalent h-GDY on Au(111) obtained with GGA and Hybrid C09 functional.


\end{document}